\documentclass[aps,pre,twocolumn,showpacs,floatfix,superscriptaddress]{revtex4}
\usepackage{amssymb}
\usepackage{amsmath}
\usepackage{graphicx}
\usepackage{latexsym}
\usepackage{amsfonts}
\usepackage{color,dsfont}

\begin{document}

\title{Staggered $\mathcal{PT}$-symmetric ladders with cubic nonlinearity}
\author{Jennie D'Ambroise}
\affiliation{Department of Mathematics and Statistics, Amherst College, Amherst, MA
01002-5000, USA}
\author{P. G. Kevrekidis}
\affiliation{Department of Mathematics and Statistics, University of Massachusetts,
Amherst, MA 01003-9305, USA}
\author{Boris A. Malomed}
\affiliation{Department of Physical Electronics, School of Electrical Engineering,
Faculty of Engineering, Tel Aviv University, Tel Aviv 69978, Israel}

\begin{abstract}
We introduce a ladder-shaped chain with each rung carrying a $\mathcal{PT}$%
-symmetric gain-loss dimer. The polarity of the dimers is \textit{staggered%
} along the chain, meaning alternation of gain-loss and loss-gain rungs.
This structure, which can be implemented as an optical waveguide array, is
the simplest one which renders the system $\mathcal{PT}$-symmetric in both
horizontal and vertical directions. The system is governed by a pair of
linearly coupled discrete nonlinear Schr\"{o}dinger (DNLS) equations with
self-focusing or defocusing cubic onsite nonlinearity. Starting from the
analytically tractable anti-continuum limit of uncoupled rungs and using the
Newton's method for continuation of the solutions with the increase of the
inter-rung coupling, we construct families of $\mathcal{PT}$-symmetric
discrete solitons and identify their stability regions. Waveforms stemming
from a single excited rung and double ones are identified. Dynamics of
unstable solitons is investigated too.
\end{abstract}

\pacs{05.45.-a, 63.20.Ry}
\maketitle

\section{Introduction}

\label{intro}

A vast research area, often called discrete nonlinear optics deals with
evanescently coupled arrayed waveguides featuring material nonlinearity \cite%
{Lederer}. Discrete arrays of optical waveguides have drawn a great deal of
interest not only because they introduce a vast phenomenology of the
nonlinear light propagation, such as e.g. the prediction \cite{vortex1} and
experimental creation \cite{vortex-exper1} of discrete vortex solitons, but
also due to the fact that they offer a unique platform for emulating the
transmission of electric signals in solid-state devices, which is obviously
interesting for both fundamental studies and applications \cite%
{Christodoulides2,Lederer}. Furthermore, the flexibility of techniques used
for the creation of virtual (photoinduced) \cite{Moti1} and permanently
written \cite{Jena1} guiding arrays enables the exploration of effects which
can be difficult to directly observe in other physical settings, such as
Anderson localization \cite{Anderson1}.

Another field in which arrays of quasi-discrete waveguides find a natural
application is the realization of the optical $\mathcal{PT}$ (parity-time)
symmetry \cite{PT1}. On the one hand, a pair of coupled nonlinear
waveguides, which carry mutually balanced gain and loss, make it possible to
realize $\mathcal{PT}$-symmetric spatial or temporal solitons (if the
waveguides are planar ones or fibers, respectively), which admit an exact
analytical solution, including their stability analysis \cite{PT-coupler}.
On the other hand, a $\mathcal{PT}$-symmetric dimer, i.e., the balanced pair
of gain and loss nodes, can be embedded, as a defect, into a regular guiding
array, with the objective to study the scattering of incident waves on the
dimer \cite{Dmitriev1,ourwork,Raymond}. We note here in passing that 
some times, also the term ``dipoles'' may be used for describing
such dimers, however we will not make use of
it here, to avoid an
overlap in terminology with classical dipoles in electrodynamics as 
discussed e.g. in~\cite{magdip}.  Discrete solitons pinned to a
nonlinear $\mathcal{PT}$-symmetric defect have been reported too \cite%
{Raymond}. Such systems, although governed by discrete nonlinear Schr\"{o}%
dinger (DNLS) equations corresponding to non-Hermitian Hamiltonians, may
generate real eigenvalue spectra (at the linear level), provided that the
gain-loss strength does not exceed a critical value, above which the $%
\mathcal{PT}$ symmetry is broken \cite{Zezyulin} [self-defocusing
nonlinearity with the local strength growing, in a one-dimensional (1D)
system, from the center to periphery at any rate faster that the distance
from the center, gives rise to stable fundamental and higher-order solitons
with \emph{unbreakable} $\mathcal{PT}$-symmetry \cite{unbreakable}].

One- and two-dimensional (1D and 2D) lattices, built of $\mathcal{PT}$
dimers, were introduced in Refs. \cite{SergeySergey1,SergeySergey2}
and \cite{Raymond-OE}, respectively. Discrete solitons, both quiescent and
moving ones, were found in these systems \cite{SergeySergey1,Raymond-OE}. In
the continuum limit, those solitons go over into those in the
above-mentioned $\mathcal{PT}$-symmetric coupler \cite{PT-coupler}.
Accordingly, a part of the soliton family is stable, and another part is
unstable. Pairs of parallel and anti-parallel coupled dimers, in the form of
$\mathcal{PT}$-symmetric plaquettes (which may be further used as building
blocks for 2D chains) were investigated too \cite{guenther,SAM}.

The objective of the present work is to introduce a \textit{staggered} chain
of $\mathcal{PT}$-symmetric dimers, with the orientations of the dimers
alternating between adjacent sites of the chain. This can also be thought of
as an extension of a plaquette from Refs. \cite{guenther,SAM} towards a
lattice. While this ladder-structured lattice is not a full 2D one, it
belongs to a class of chain systems which may be considered as $1.5$D models~%
\cite{vakakis}.

As shown in Section II, where the model is introduced, the fundamental
difference from the previously studied ones is the fact that such a system,
although being nearly one-dimensional, actually realizes the $\mathcal{PT}$
symmetry in the 2D form, with respect to both horizontal and vertical
directions. In Section III we start the analysis from the solvable
anti-continuum limit (ACL) \cite{Panos}, in which the rungs of the ladder
are uncoupled (in the opposite continuum limit, the ladder degenerates into
a single NLS equation). Using parametric continuation from this limit makes
it possible to construct families of discrete solitons in a numerical form.
Such solution branches are initiated, in the ACL, by a single excited rung,
as well as by the excitation confined to several rungs. The soliton
stability is systematically analyzed in Section III too and, if the modes
are identified as unstable, their evolution is examined to observe the
instability development. The paper is concluded by Section IV, where also
some directions for future study are presented.

\section{The model}

\label{model}

\begin{figure}[tbp]
\begin{center}
\includegraphics[width=8cm,angle=0,clip]{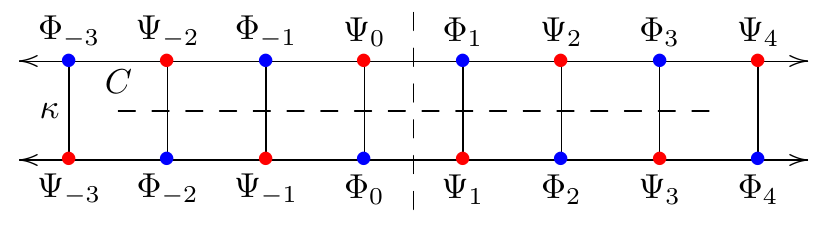}
\end{center}
\caption{(Color online) The staggered ladder-shaped lattice with horizontal (%
$C$) and vertical ($\protect\kappa $) coupling constants. Red and blue dots
designate sites carrying the mutually balanced gain and loss, respectively.
The dashed lines designate the horizontal and vertical axes of the $\mathcal{%
PT}$ symmetry.}
\label{xypic}
\end{figure}

We consider the ladder configuration governed by the DNLS system with
intersite coupling constant $C$,
\begin{gather}
i\frac{d\Psi _{n}}{dt}+\frac{C}{2}\left( \Phi _{n+1}+\Phi _{n-1}-2\Psi
_{n}\right)  \notag \\
+\sigma |\Psi _{n}|^{2}\Psi _{n}=i\gamma \Psi _{n}-\kappa \Phi _{n},
\label{tdnls} \\
i\frac{d\Phi _{n}}{dt}+\frac{C}{2}\left( \Psi _{n+1}+\Psi _{n-1}-2\Phi
_{n}\right)  \notag \\
+\sigma |\Phi _{n}|^{2}\Phi _{n}=-i\gamma \Phi _{n}-\kappa \Psi _{n},  \notag
\end{gather}%
where evolution variable $t$ is the propagation distance, in terms of the
optical realization. Coefficients $+i\gamma $ and $-i\gamma $ with $\gamma
>0 $ represent $\mathcal{PT}$-symmetric gain-loss dimers, whose orientation
is staggered (alternates) along the ladder, the sites carrying gain and loss
being represented by amplitudes $\Psi _{n}(t)$ and $\Phi _{n}(t)$,
respectively. Cubic nonlinearity with coefficient $\sigma $ is present at
every site, and $\kappa >0$ accounts for the vertical coupling along the
ladder's rungs, each representing a $\mathcal{PT}$-symmetric dimer. The
system is displayed in Fig. \ref{xypic}. As seen in the figure, the nearly
1D ladder realizes the $\mathcal{PT}$ symmetry in the 2D form, with respect
to the horizontal axis, running between the top and bottom strings, and,
simultaneously, with respect to any vertical axis drawn between adjacent
rungs.

By means of obvious rescaling, we can fix $|\sigma |=1$, hence the
nonlinearity coefficient takes only two distinct values, which correspond,
respectively, to the self-focusing and defocusing onsite nonlinearity, $%
\sigma =+1$ and $\sigma =-1$. The usual DNLS equation admits the sign
reversal of $\sigma $ by means of the well-known staggering transformation
\cite{Panos}. However, once we fix $\gamma >0$ (and also $\kappa >0$) in Eq.
(\ref{tdnls}), this transformation cannot be applied, as it would also
invert the signs of $\gamma $ and $\kappa $.

The single self-consistent continuum limit of system (\ref{tdnls}),
corresponding to $C\rightarrow \infty $, is possible for the 
fields related by $\Phi =e^{i\delta }\Psi $, with phase shift $\delta
=\gamma /C$. Replacing, in this limit, the finite-difference derivative by
the one with respect to the continuous coordinate, $x\equiv n/\sqrt{C}$,
yields the standard NLS equation,
\begin{equation}
i\frac{\partial \Psi }{\partial t}+\frac{1}{2}\frac{\partial ^{2}\Psi }{%
\partial x^{2}}+\sigma |\Psi |^{2}\Psi =-\kappa \Psi .
\end{equation}%
Given its ``standard'' nature, leading to a full mutual cancellation
of the gain and loss terms, we will not pursue this limit further.
Instead, as shown below, 
we will use as a natural starting point for examining nontrivial localized
modes in the discrete system (\ref{tdnls}) the opposite ACL, which
corresponds to $C\rightarrow 0$, i.e., the set of uncoupled rungs.

Stationary solutions to Eqs. (\ref{tdnls}) with real propagation constant $%
\Lambda $ are sought in the usual form, $\Psi _{n}=e^{i\Lambda t}u_{n}$ and $%
\Phi _{n}=e^{i\Lambda t}v_{n}$, where functions $u_{n}$ and $v_{n}$ obey the
stationary equations,
\begin{gather}
-\Lambda u_{n}+\frac{C}{2}\left( v_{n+1}+v_{n-1}-2u_{n}\right)  \notag \\
+\sigma |u_{n}|^{2}u_{n}=i\gamma u_{n}-\kappa v_{n},  \label{stat} \\
-\Lambda v_{n}+\frac{C}{2}\left( u_{n+1}+u_{n-1}-2v_{n}\right)  \notag \\
+\sigma |v_{n}|^{2}v_{n}=-i\gamma v_{n}-\kappa u_{n}.  \notag
\end{gather}%
Numerical solutions of these equations for discrete solitons are produced in
the next section. To analyze the stability of the solutions, we add
perturbations with an infinitesimal amplitude $\varepsilon $ and frequencies
$\omega $,
\begin{eqnarray}
\Psi _{n}(t) &=&(u_{n}+\varepsilon (a_{n}e^{i\omega t}+b_{n}e^{-i\omega ^{\ast
}t}))e^{i\Lambda t},  \notag \\
\Phi _{n}(t) &=&(v_{n}+\varepsilon (c_{n}e^{i\omega t}+d_{n}e^{-i\omega ^{\ast
}t}))e^{i\Lambda t}.  \label{pert}
\end{eqnarray}%
The linearization of Eq. (\ref{tdnls}) with respect to the small
perturbations leads to the eigenvalue problem,
\begin{equation}
M\left[
\begin{array}{c}
a_{n} \\
b_{n}^{\ast } \\
c_{n} \\
d_{n}^{\ast }%
\end{array}%
\right] =\omega \left[
\begin{array}{c}
a_{n} \\
b_{n}^{\ast } \\
c_{n} \\
d_{n}^{\ast }%
\end{array}%
\right] ,  \label{lineq}
\end{equation}%
where $M$ is a $4N\times 4N$ matrix for the ladder of length $N$. Using
standard indexing, $N\times N$ submatrices of $M$ are defined as
\begin{eqnarray}
M_{11} &=& \mathrm{diag}(p_n^*-\Lambda-C), \notag \\
M_{22} &=& \mathrm{diag}(\Lambda+C-p_n), \notag \\
M_{33} &=& \mathrm{diag}(q_n-\Lambda-C), \notag \\
M_{44} &=& \mathrm{diag}(\Lambda+C-q_n^*), \notag \\
M_{12}=-M_{21}^{\ast } &=&\mathrm{diag}(\sigma u_{n}^{2}),  \label{M} \\
M_{34}=-M_{43}^{\ast } &=&\mathrm{diag}(\sigma v_{n}^{2}),  \notag \\
M_{13}=M_{31} &=&-M_{24}=-M_{42}=\frac{C}{2}G+\mathrm{diag}(\kappa ),  \notag
\end{eqnarray}
\begin{eqnarray}
p_{n} &\equiv & i\gamma +2\sigma |u_{n}|^{2},  \notag \\
q_{n} &\equiv & i\gamma +2\sigma |v_{n}|^{2},  \label{pq}
\end{eqnarray}
where $G$ is an $N\times N$ matrix of zero elements, except for the super-
and sub-diagonals that contain all ones.

For the zero solution of the stationary equation \eqref{stat}, $%
u_{n}=v_{n}=0,~$matrix $M$ has constant coefficients, hence perturbation
eigenmodes can be sought for as $%
a_{n}=Ae^{ikn},b_{n}=0,c_{n}=Be^{ikn},d_{n}=0$. Then Eq. \eqref{lineq}
becomes a $2\times 2$ system, whose eigenvalues can be found explicitly:
\begin{equation}
\omega =-(\Lambda +C)\pm \sqrt{(C\cos k+\kappa )^{2}-\gamma ^{2}},
\label{zernu}
\end{equation}%
so that $\omega $ is real only for $C\leq \kappa -\gamma $. In other words, the
$\mathcal{PT}$-symmetry is broken, with $i\omega $ acquiring a positive real
part, which drives the exponential growth of the perturbations, at
\begin{equation}
\gamma >\gamma _{\mathrm{cr}}^{(1)}(C)\equiv \kappa -C.  \label{cr}
\end{equation}%
It is interesting to observe here that the coupling between the rungs
decreases the size of the interval of the unbroken $\mathcal{PT}$-symmetry
of the single dimer~\cite{PT1,Zezyulin}.

In the stability region, Eq. (\ref{zernu}) demonstrates that real
perturbation frequencies take values in the following intervals:
\begin{gather}
-(\Lambda +C)-\sqrt{(\kappa +C)^{2}-\gamma ^{2}}<\omega <  \notag \\
-(\Lambda +C)-\sqrt{(\kappa -C)^{2}-\gamma ^{2}},  \notag \\
-(\Lambda +C)+\sqrt{(\kappa -C)^{2}-\gamma ^{2}}<\omega <  \notag \\
-(\Lambda +C)+\sqrt{(\kappa +C)^{2}-\gamma ^{2}}.  \label{intervals}
\end{gather}%
Similarly, for the perturbations in the form of $%
a_{n}=0,b_{n}=Ae^{ikn},c_{n}=0,d_{n}=Be^{ikn}$ the negatives of expressions (%
\ref{zernu}) are also eigenvalues of the zero stationary solution, and at\ $%
\gamma <\gamma _{\mathrm{cr}}^{(1)}(C)$, they fall into the negatives of
intervals (\ref{intervals}). 

Simultaneously, Eq. (\ref{zernu}) and its negative counterpart give the
dispersion relation for plane waves (``phonons") in the linearized version
of Eq. (\ref{tdnls}). Accordingly, intervals (\ref{intervals}), along with
their negative counterparts, represent phonon bands of the linearized system.

In Section \ref{sol} we produce stationary solutions in the form of discrete
solitons. This computation begins by finding exact solutions for the ACL, $%
C=0$, and then the continuing the solutions numerically to $C>0$, by means
of the Newton's method for each $C$ (i.e., utilizing the converging solution
obtained for a previous value of $C$ as an initial seed for the Newton's
algorithm with $C\rightarrow C+\Delta C$). As suggested by Eq. (\ref{cr}),
we restrict the analysis to $0<\gamma \leq \kappa $, so as to remain within
the $\mathcal{PT}$-symmetric region at $C=0$. Subsequently, the stability
interval of the so constructed solutions is identified, in a numerical form
too. %Then, in Section \ref%
%{Cneq0} it will be shown that, for increasing $C$, the Newton's method
%indeed does not converge to stable solutions in region (\ref{cr}).

\section{Discrete solitons and their stability}

\label{sol}

\subsection{The anti-continuum limit (ACL), $C=0$}

\label{solC=0}

To construct stationary localized solutions of Eqs. (\ref{tdnls}) at $C=0$,
when individual rungs are decoupled, we substitute
\begin{equation}
u_{n}=e^{i\delta _{n}}v_{n}  \label{rung-ansatz}
\end{equation}
with real $\delta _{n}$ in Eq. (\ref{stat}), which yields relations
\begin{equation}
\gamma =-\kappa \sin \delta _{n},~~\sigma |v_{n}|^{2}=-\kappa \cos \delta
_{n}+\Lambda .  \label{rung_rel}
\end{equation}
For the uncoupled ladder, one can specify either a
single-rung solution, with fields at all sites set equal to zero except for $%
u_{1}$ satisfying Eq. (\ref{rung_rel}), or a double-rung solution with
nonzero fields $u_{1}$ and $u_{2}$ satisfying the same equations. We focus
on these two possibilities in the ACL (although larger-size solutions are
obviously possible too). These are the direct counterparts of the
single-node and two-node solutions that have been extensively studied in 1D
and 2D DNLS models~\cite{Panos}.

We take parameters satisfying constraints
\begin{equation}
\sigma >0,~\Lambda >\kappa ,  \label{par}
\end{equation}%
to make the second equation (\ref{rung_rel}) self-consistent. Then, two
solution branches for $\delta _{n}$ are possible. The first branch satisfies
$-\pi /2\leq \delta _{\mathrm{in}}\equiv \arcsin (-\gamma /\kappa )\leq 0$
and $\cos (\delta _{\mathrm{in}})\geq 0$. Choosing a solution with $\delta
_{n}=\delta _{\mathrm{in}}$ in the rung carrying nonzero fields, we name it
an \textit{in-phase rung}, as the phase shift between the gain and loss
poles of the respective dimer is smaller than $\pi /2$, namely, $|\arg
(vu^{\ast })|\in \lbrack 0,\pi /2]$. The second branch satisfies $-\pi \leq
\delta _{\mathrm{out}}\equiv -\pi +|\delta _{\mathrm{in}}|\leq -\pi /2$ and $%
\cos \delta _{\mathrm{out}}\leq 0$. The rung carrying the solution with $%
\delta _{n}=\delta _{\mathrm{out}}$ is called an \textit{out-of-phase }one,
as the respective phase shift between the elements exceeds $\pi /2$, \textit{%
viz}., $|\arg (vu^{\ast })|\in \lbrack \pi /2,\pi ]$. The two branches meet
and disappear at $\gamma =\kappa $, when $\delta _{\mathrm{in}}=\delta _{%
\mathrm{out}}=-\pi /2$. Recall that $\gamma =\kappa =\gamma _{\mathrm{cr}%
}^{(1)}(C=0)$ [see Eq. (\ref{cr})] is the boundary of the $\mathcal{PT}$%
-symmetric region for $C=0$. These branches can be also be considered as
stemming from the Hamiltonian limit of $\gamma =0$, where $\delta _{\mathrm{%
in}}=0$ and $\delta _{\mathrm{out}}=\pi $ correspond, respectively, to the
usual definitions of the in- and out-of-phase Hamiltonian dimers.

The stability eigenfrequencies for stationary solitons at $C=0$ can be
readily calculated analytically in the ACL~\cite{Zezyulin}. In this case, $M$
has the same eigenvalues as submatrices
\begin{equation*}
m_{0}=\left(
\begin{array}{cccc}
-\Lambda -i\gamma & 0 & \kappa & 0 \\
0 & \Lambda -i\gamma & 0 & -\kappa \\
\kappa & 0 & -\Lambda +i\gamma & 0 \\
0 & -\kappa & 0 & \Lambda +i\gamma%
\end{array}%
\right) ,
\end{equation*}%
which is associated with zero-amplitude (unexcited) rungs, and
\begin{equation*}
m_{n}=\left(
\begin{array}{cccc}
-\Lambda + p_{n}^{\ast } & \sigma u_{n}^{2} & \kappa & 0 \\
-\sigma (u_{n}^{\ast })^{2} & \Lambda - p_{n} & 0 & -\kappa \\
\kappa & 0 & -\Lambda + q_{n} & \sigma v_{n}^{2} \\
0 & -\kappa & -\sigma (v_{n}^{\ast })^{2} & \Lambda - q_{n}^* %
\end{array}%
\right) ,
\end{equation*}%
associated with the excited ones, which carry nonzero stationary fields,
with $v_{n}$, $u_{n}$ taken as per Eqs. (\ref{rung-ansatz}) and (\ref%
{rung_rel}). In other words, each of the four eigenvalues of $m_{0}$,
\begin{equation}
\omega =\pm \Lambda \pm \sqrt{\kappa ^{2}-\gamma ^{2}},  \label{nexeig}
\end{equation}%
is an eigenvalue of $M$ with multiplicity equal to the number of
zero-amplitude rungs, while each of the four eigenvalues of $m_{n}$,
\begin{equation}
\omega =\pm 0,~\pm 2\Lambda \sqrt{2\alpha _{\ast }^{2}-\alpha _{\ast }},
\label{exeig}
\end{equation}%
appears as an eigenvalue of $M$ with multiplicity equal to the number of
excited rungs. Here $\alpha _{\ast }=\alpha _{\mathrm{in}}=\kappa \cos
(\delta _{\mathrm{in}})/\Lambda \equiv \sqrt{(\kappa ^{2}-\gamma
^{2})/\Lambda ^{2}}$, and $\alpha _{\ast }=\alpha _{\mathrm{out}}=\kappa
\cos (\delta _{\mathrm{out}})/\Lambda \equiv -\sqrt{(\kappa ^{2}-\gamma
^{2})/\Lambda ^{2}}$ for an in- and out-of-phase rung, respectively.

Equation (\ref{exeig}) shows that the out-of-phase excited rung is always
stable, as it has $\mathrm{Re}(i\omega )=0$. Similarly, the in-phase excited
rung is stable for $\kappa ^{2}-\gamma ^{2}\geq {\Lambda ^{2}}/{4}$, %or $%
%\kappa =\gamma $,
and unstable for $0<\kappa ^{2}-\gamma ^{2}<\Lambda ^{2}/4$. Thus, for
solutions that contain an excited in-phase rung in the initial configuration
at $C=0$, there are the two critical values, \textit{viz}., $\gamma _{%
\mathrm{cr}}^{(1)}(C=0)=\kappa $ given by Eq. (\ref{cr}), and the additional
one, which designates the instability area for the uncoupled in-phase rungs:
\begin{equation}
\gamma >\gamma _{\mathrm{cr}}^{(2)}(C=0)=\sqrt{\kappa ^{2}-\Lambda ^{2}/4}.
\label{cr2}
\end{equation}

A choice alternative to Eq. (\ref{par}) is
\begin{equation}
\sigma <0,~\Lambda <-\kappa .  \label{pars=-1}
\end{equation}%
In this case, the analysis differs only in that the sign of $\alpha _{\ast }$
in Eq. \eqref{exeig} is switched. That is, the in-phase rung is now
associated to negative $\alpha _{\ast }=\alpha _{\mathrm{in}}=\kappa \cos
(\delta _{\mathrm{in}})/\Lambda =-\sqrt{(\kappa ^{2}-\gamma ^{2})/\Lambda
^{2}}$, while the out-of-phase one to positive $\alpha _{\ast }=\alpha _{\mathrm{%
out}}=\kappa \cos (\delta _{\mathrm{out}})/\Lambda \equiv \sqrt{(\kappa
^{2}-\gamma ^{2})/\Lambda ^{2}}$. In this case, the in-phase rung is always
stable, while its out-of-phase counterpart is unstable at $\gamma >\gamma _{%
\mathrm{cr}}^{(2)}(C=0)$, see Eq. (\ref{cr2}).

\subsection{Discrete solitons at $C>0$}

\label{Cneq0}

To construct soliton solutions for coupling constant $C$ increasing in steps
of $\Delta C$, we write Eq. (\ref{stat}) as a system of $4N$ equations for $%
4N$ real unknowns $w_{n},x_{n},y_{n},z_{n}$, with $u_{n}\equiv w_{n}+ix_{n}$%
, $v_{n}\equiv y_{n}+iz_{n}$. Then we apply the Newton's method with the
initial guess at each step taken as the soliton solution found at the
previous value of $C$, as mentioned above. Thus, the initial guess at $%
C=\Delta C$ is the analytical solution for $C=0$ given by Eqs. (\ref%
{rung-ansatz})-(\ref{rung_rel}) with parameters taken according to either
Eq. \eqref{par} or Eq. \eqref{pars=-1}.

\begin{figure}[tbp]
\centerline{\includegraphics[scale=.55]{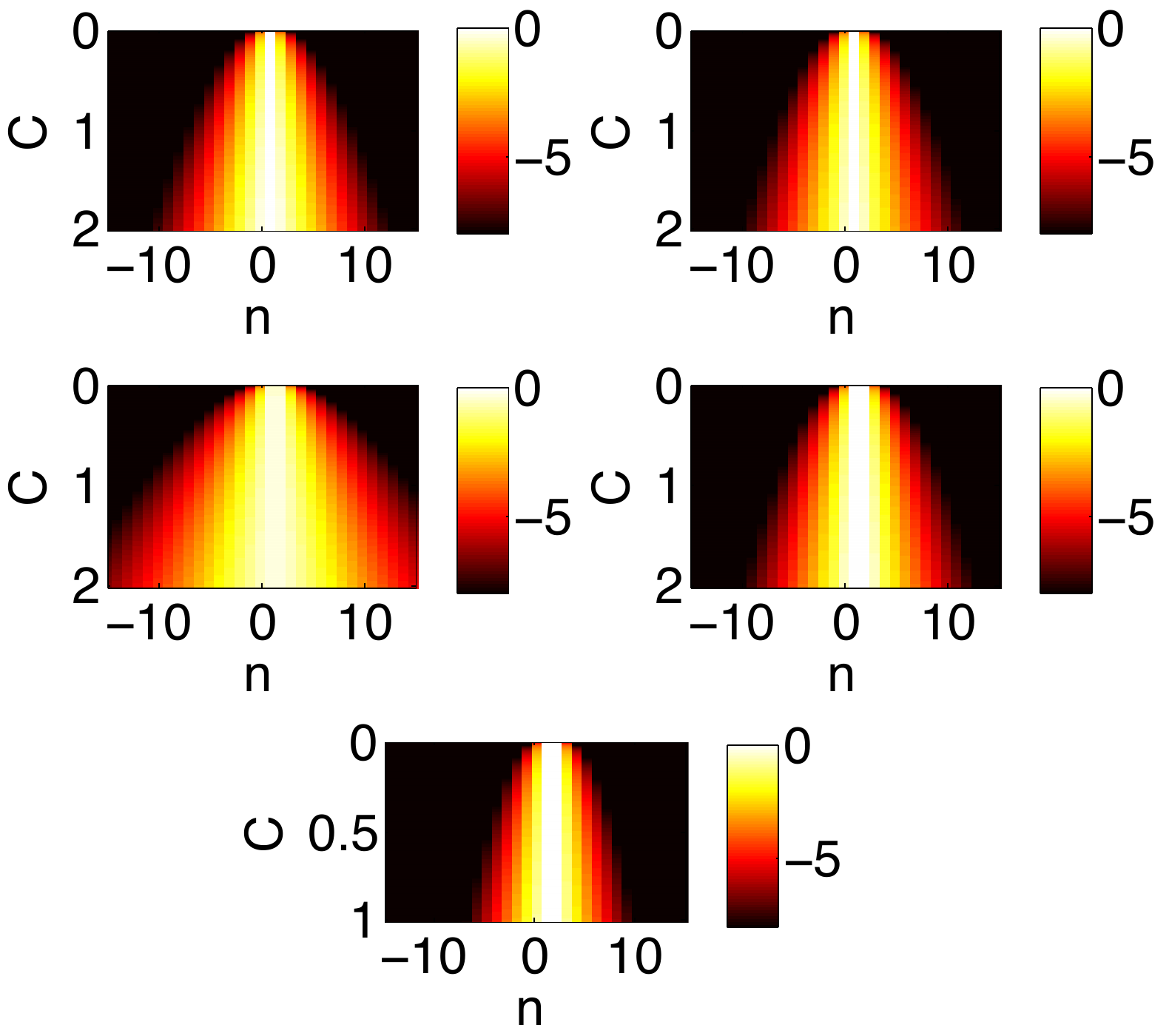} }
\caption{(Color online) Plots of $\log_{10}(|u_{n}|^{2})$, where $u_{n}$ at $C=0$
is given by Eqs. (\protect\ref{rung-ansatz}) and (\protect\ref{rung_rel}),
and at $C>0$ the soliton solutions $u_{n}$ are obtained by the continuation
in $C$, see the text. Common parameters are $\protect\gamma =1$, $\protect%
\kappa =1.9$ and $\protect\sigma =1$. The initial configuration of the
excited rungs at $C=0$ and parameters are: a single in-phase rung with $%
\protect\delta _{1}=\protect\delta _{\mathrm{in}}$, $\Lambda
=2.5,N=40,\Delta C=0.001$ (top left), a single out-of-phase rung with $%
\protect\delta _{1}=\protect\delta _{\mathrm{out}}$, $\Lambda
=2.5,N=40,\Delta C=0.001$ (top right), two in-phase rungs with $\protect%
\delta _{1}=\protect\delta _{2}=\protect\delta _{\mathrm{in}}$, $\Lambda
=2,N=80,\Delta C=0.001$ (middle left), two out-of-phase rungs with $\protect%
\delta _{1}=\protect\delta _{2}=\protect\delta _{\mathrm{out}}$, $\Lambda
=2.5,N=80,\Delta C=0.001$ (middle right), and, finally, a mixed state
carried by two rungs with $\protect\delta _{1}=\protect\delta _{\mathrm{in}}$%
, $\protect\delta _{2}=\protect\delta _{\mathrm{out}}$, $\Lambda
=2.5,N=80,\Delta C=0.00001$ (bottom center). Plots of $\log_{10} (|v_{n}|^{2})$
are identical to those of $\log_{10} \left( |u_{n}|^{2}\right) $. As $C$
increases, small amplitudes appear at adjacent rungs, and the soliton gains
width. The corresponding second moment, $w(C)$, defined as per Eq. \eqref{w}%
, is shown in Fig. \protect\ref{s=1w}. The stability of the solitons shown
here is predicted by eigenvalue plots displayed in Fig. \protect\ref{s=1mxRE}
for $\protect\gamma =1$.}
\label{s=1abs}
\end{figure}

\begin{figure}[tbp]
\centerline{\includegraphics[scale=.5]{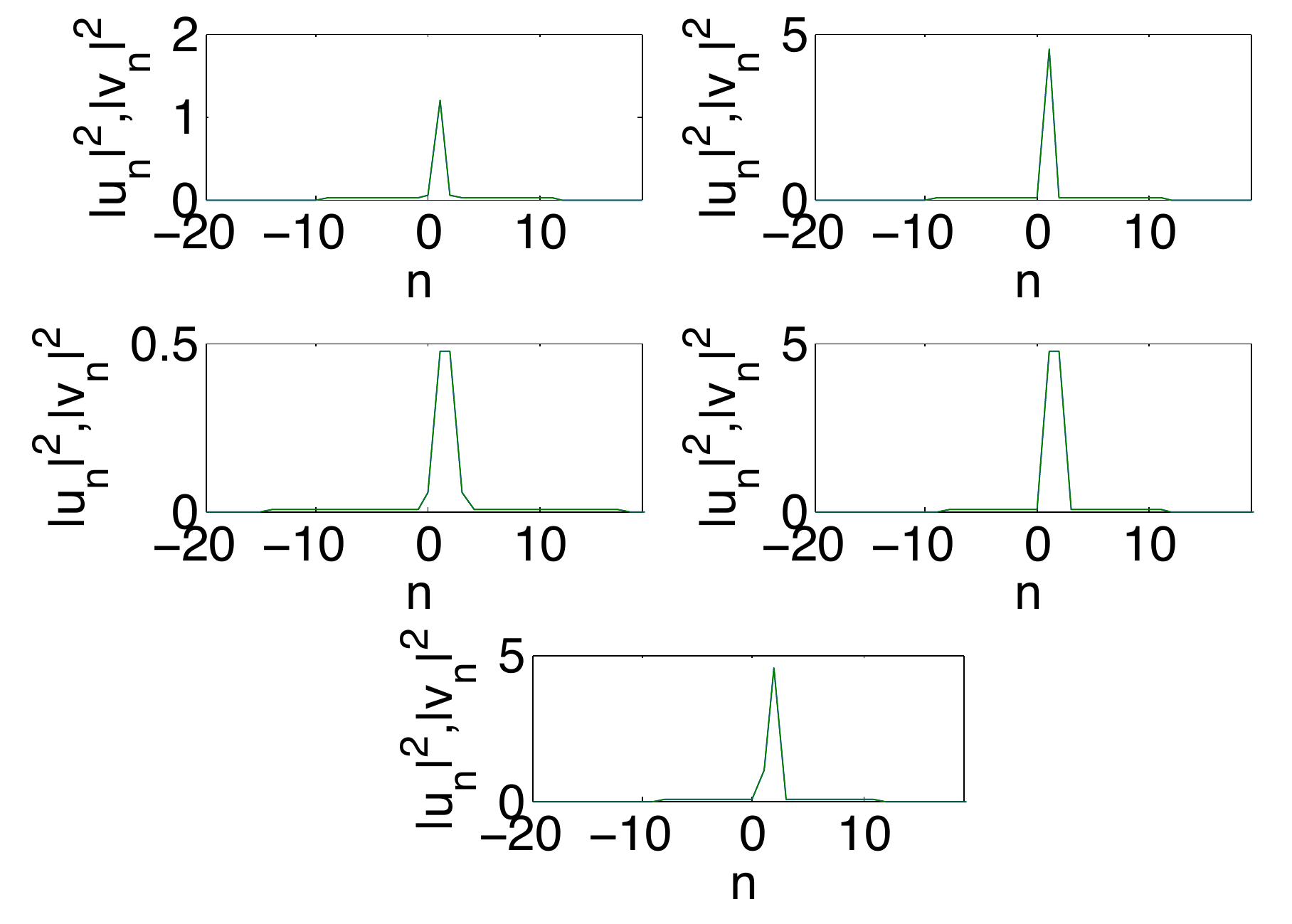} }
\caption{Profiles of discrete solitons for $C=0.4$. The configurations of
the initial ($C=0$) solution and other parameters follow the same pattern as
in Fig. \protect\ref{s=1abs}. }
\label{s=1prof}
\end{figure}

\begin{figure}[tbp]
\centerline{\includegraphics[scale=.55]{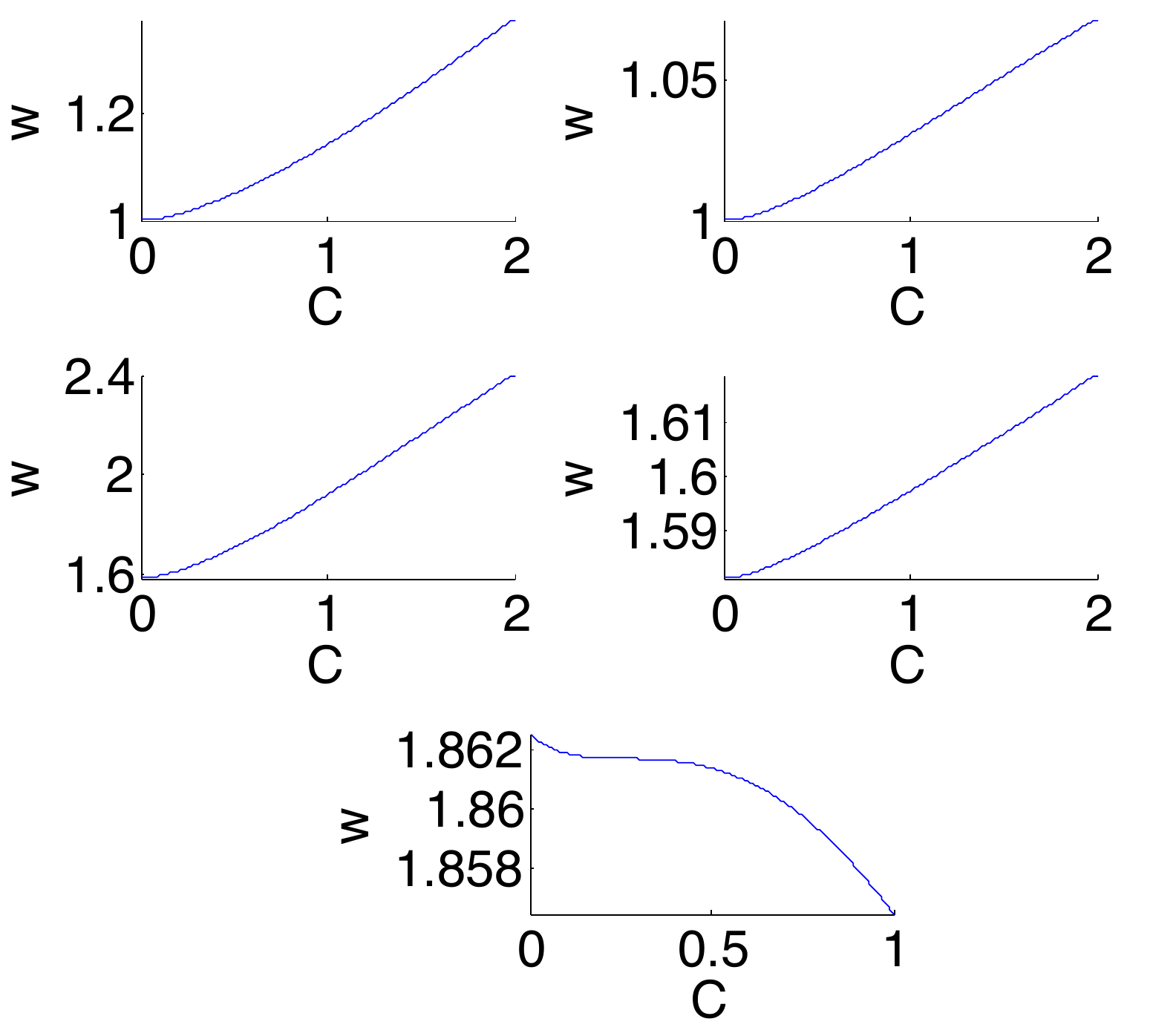} }
\caption{(Color online) Width diagnostic $w$, defined as per Eq. (\protect
\ref{w}), corresponding to each of the plots in Fig. \protect\ref{s=1abs}. }
\label{s=1w}
\end{figure}

Figure \ref{s=1abs} shows $|u_{n}|^{2}$ for the solutions identified by this
process on a (base $10$) logarithmic scale as a function of $C$ for parameters taken as
per Eq. \eqref{par}. The logarithmic scale is chosen, as it
yields a clearer picture of the variation of the solution's spatial width,
as $C$ varies. The different solutions displayed in Fig. \ref{s=1abs}
include those seeded by the single excited in- and out-of-phase rungs (the
top row), and two-rung excitations for which there are three possibilities:
in- in both and out-of-phase structures in both rungs (the second row), as
well as a mixed structure involving one rung initially excited in-phase, and
the other one excited out-of-phase (the bottom row). In Figure \ref{s=1prof}
we plot $|u_{n}|^{2}$ for fixed $C$ across the various configurations. A
point that is clearly illustrated by this figure, which is not evident on
the logarithmic scale of Fig. \ref{s=1abs}, is the 
asymmetric spatial structure
of the mixed-phase solution of the bottom row (asymmetric solitary waves
have also been proposed
in full 2D lattices \cite{Asymm}; see also
the detailed analysis of~\cite{Panos}). Equations (\ref{rung-ansatz})-(\ref%
{rung_rel}) show that, for $C=0$, since the out-of-phase case corresponds to
$\cos (\delta _{\mathrm{out}})\leq 0$, the amplitude of the out-of-phase
rung, $|v_{n}|=|u_{n}|$, with $\sigma =+1$, is larger in comparison with its
in-phase counterpart, which has $\cos (\delta _{\mathrm{in}})\geq 0$. The
asymmetry for the mixed-phase solution persists for $C>0$, Fig. \ref{s=1prof}
showing an example of this. Then, Fig. \ref{s=1w} shows more explicitly the
increasing width of the soliton, using the second moment of the density
distribution, as the respective diagnostic,
\begin{equation}
w(C)\equiv \sqrt{\frac{\displaystyle\sum_{n}n^{2}|u_{n}|^{2}}{\displaystyle%
\sum_{n}|u_{n}|^{2}}},  \label{w}
\end{equation}%
versus $C$ for the solutions shown in Fig. \ref{s=1abs}. It is relevant to
point out that the variation of this width-measuring quantity is fairly weak
in the case of the out-of-phase solutions and mixed ones, while it is more
significant in the case of the single and double in-phase excited rungs.

%{\Huge [In Fig. \ref{s=1abs} the solitons' shape seems symmetric, while in
%Fig. \ref{s=1prof} the shapes are obviously asymmetric. This point should
%definitely be discussed.]}

\begin{figure}[tbp]
\centerline{\includegraphics[scale=.55]{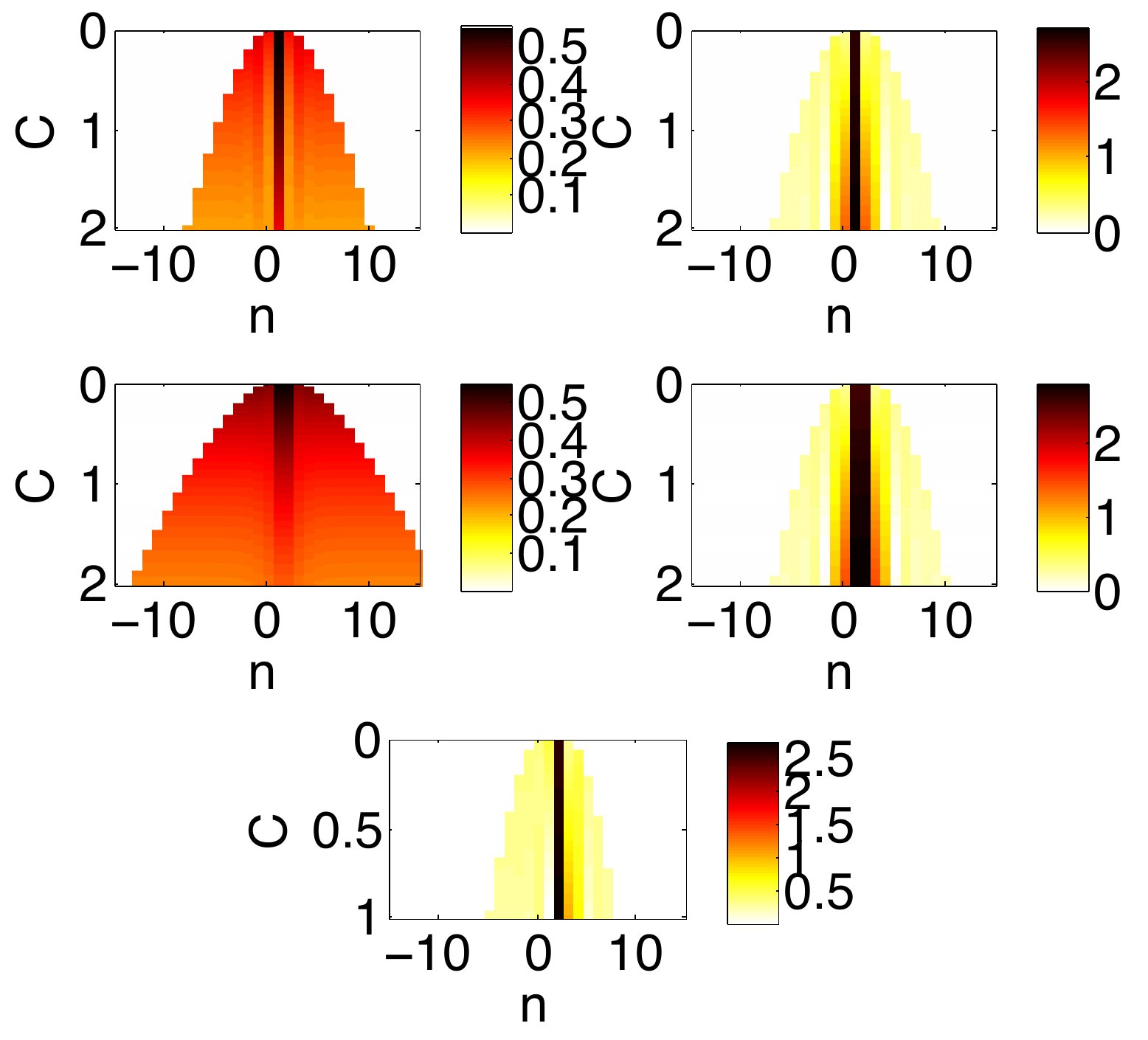} }
\caption{The phase shift between two edges of the rungs, $|\mathrm{arg}%
(v_{n}u_{n}^{\ast })|$, plotted as a function of $C$ and $n$, where $u$ is
the solution whose absolute value is presented in Fig. \protect\ref{s=1abs}.
The value of $|\mathrm{arg}(v_{n}u_{n}^{\ast })|$ is set to zero for any $n$
at which $\log_{10} \left( |u|^{2}\right) \leq -6$ in Fig. \protect\ref{s=1abs}.
In other words, the phase shift is shown as equal to zero when the amplitude
is too small. For the top left and middle left plots, the soliton's field is
different from zero at one or two in-phase rung(s) when $C=0$, and as $C$
increases the solutions stay in-phase. For the top right and middle right
plots, the field at $C=0$ is nonzero and out-of-phase at the one or two
central rungs, and, as $C$ increases, the fields at these rungs, and at two
rungs on either side of the central ones, tend to be out-of-phase, while the
field at other rungs, located farther away, tend to be in-phase. Similarly,
in the bottom plot, where at $C=0$ the $n=1$ rung is in-phase and the $n=2$
one is out-of-phase, as $C$ increases, most rungs tend to be in-phase,
except for $n=2,3$. }
\label{s=1arg}
\end{figure}

\begin{figure}[tbp]
\centerline{\includegraphics[scale=.55]{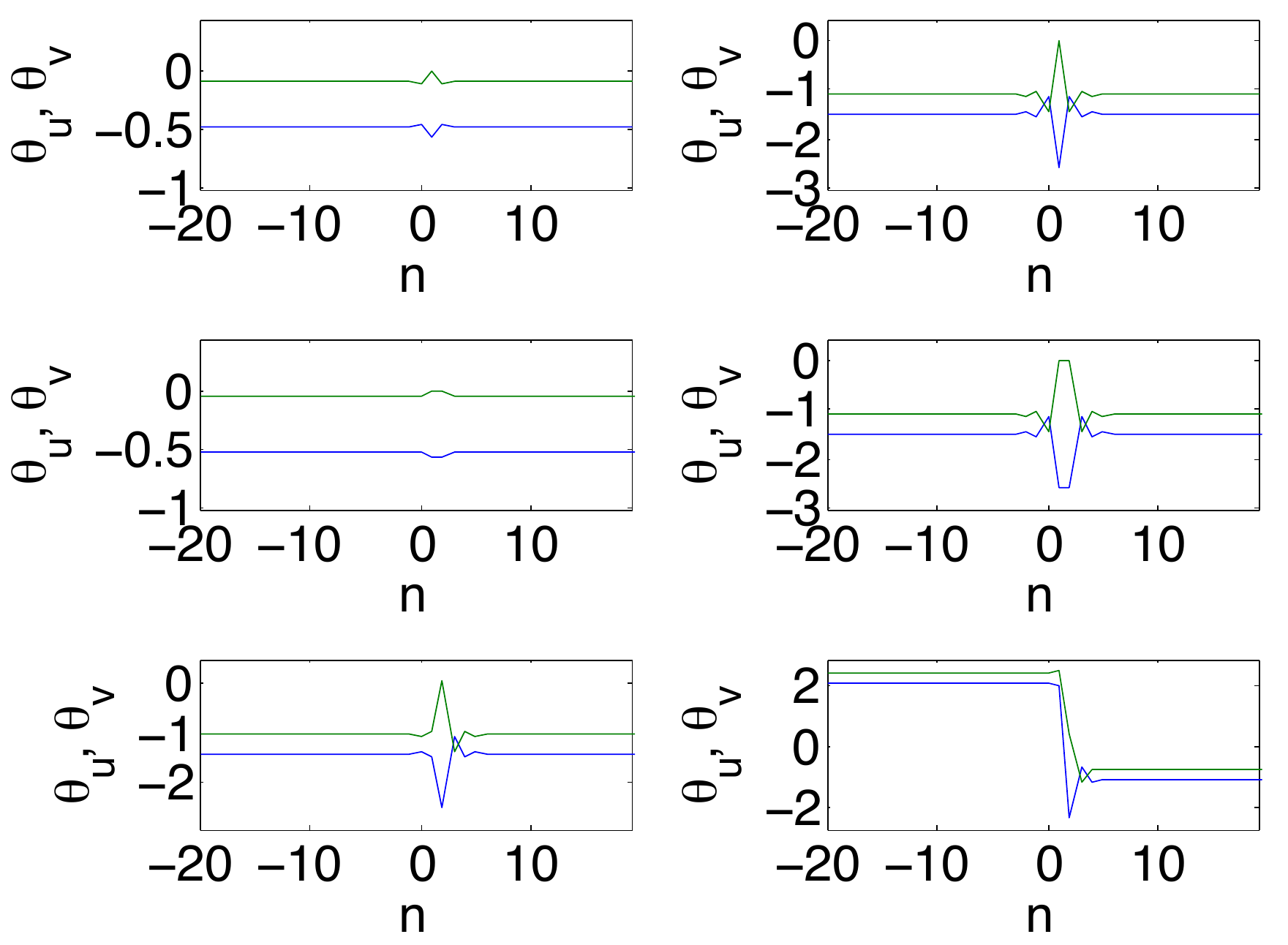} }
\caption{The phase of each rung, $\theta_u := \mathrm{arg}(u_n)$ (blue) and
$\theta_v := \mathrm{arg}(v_n)$ (green) with values in the interval $(-\pi,\pi]$,
plotted as a function of the spatial variable $n$.  The top four plots and bottom-most left plot have 
fixed parameter values $C = 0.4, \gamma = 1$ and $\sigma = 1$.  That is, the $u_n, v_n$ 
solutions for these five plots are the same as those whose modulus is plotted in Fig. \ref{s=1prof}.
In the bottom right plot we show $\theta_u, \theta_v$ for the same fixed parameter values as the bottom left plot except with $\gamma = 0.8$.  The bottom two plots show the different phase profiles can that arise 
for different $\gamma$ values, with either same (left) or opposite (right) signs for the phase on the outer portions 
of the ladder.  
 }
\label{s=1arguv}
\end{figure}

In Fig. \ref{s=1arg} the absolute value of the phase difference between
fields $u_{n}$ and $v_{n}$ at two sides of the ladder is shown. In other
words, this figure shows whether each rung of the ladder belongs to the
in-phase or out-of-phase type, as a function of $C$. This figure reveals
that, as $C$ increases, there is a progressive spatial expansion (across $n$) 
in the number of sites supporting a phase difference that develops around
the initially excited sites.  The individual phases of $u_n$ and $v_n$ are shown 
in Fig. \ref{s=1arguv}.  We show in the bottom two plots of Fig. \ref{s=1arguv} that 
two different types of phase profiles can arise; one type has phase with the same sign 
on both the left and right sides of the outer portion of the ladder, and the second type has
 phases that are of opposite sign on the left and right sides of the outer portion of the ladder.
 We address this point more in the next section where we discuss stability.

\begin{figure}[tbp]
\centerline{\includegraphics[scale=.55]{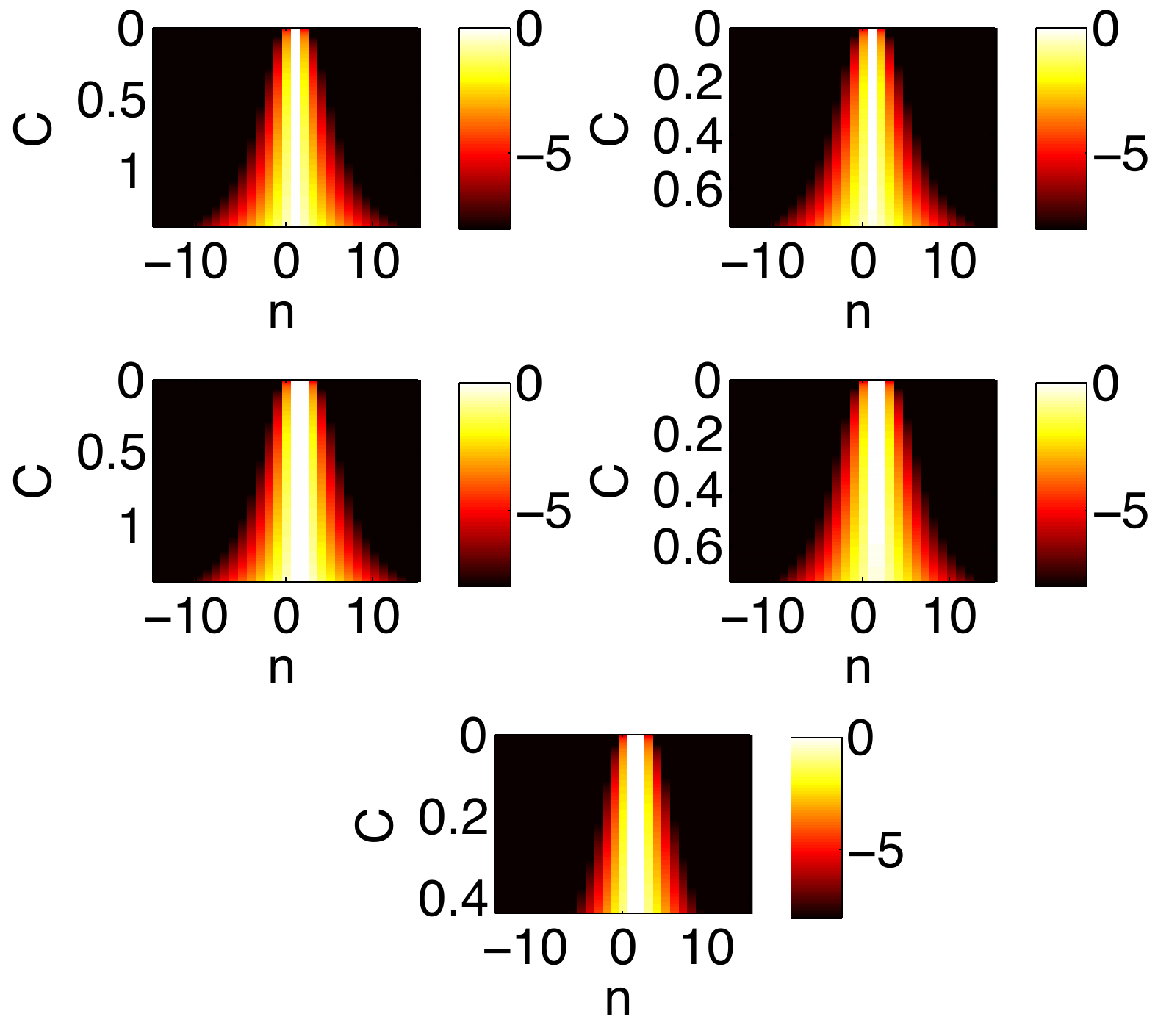} }
\caption{(Color online) The same as Fig. \protect\ref{s=1abs}, but for $%
\protect\sigma =-1$. Common parameters are $\protect\kappa =2$ and $\protect%
\gamma =1$. The initial values of $\protect\delta _{1}$ and $\protect\delta %
_{2}$ at $C=0$ follow the pattern of Fig. \protect\ref{s=1abs}. Other
parameters are: $\Lambda =5, N=80, \Delta C=0.001$ on the top left, $\Lambda
=3.5, N=80, \Delta C=0.001$ on the top right, $\Lambda =5, N=80, \Delta
C=0.001$ on the middle left, $\Lambda =3.5, N=40, \Delta C=0.001$ on the
middle right, and finally $\Lambda =3.085,N=40,\Delta C=0.0001$ on the
bottom center. As $C $ increases, small amplitudes appear at adjacent sites,
and the soliton gains width, as shown by means of $w$ in Fig. \protect\ref%
{s=-1w}. The stability of the solitons shown here is predicted by the
eigenvalue plots in Fig. \protect\ref{s=-1mxRE} at $\protect\gamma =1$. }
\label{s=-1abs}
\end{figure}

\begin{figure}[tbp]
\centerline{\includegraphics[scale=.5]{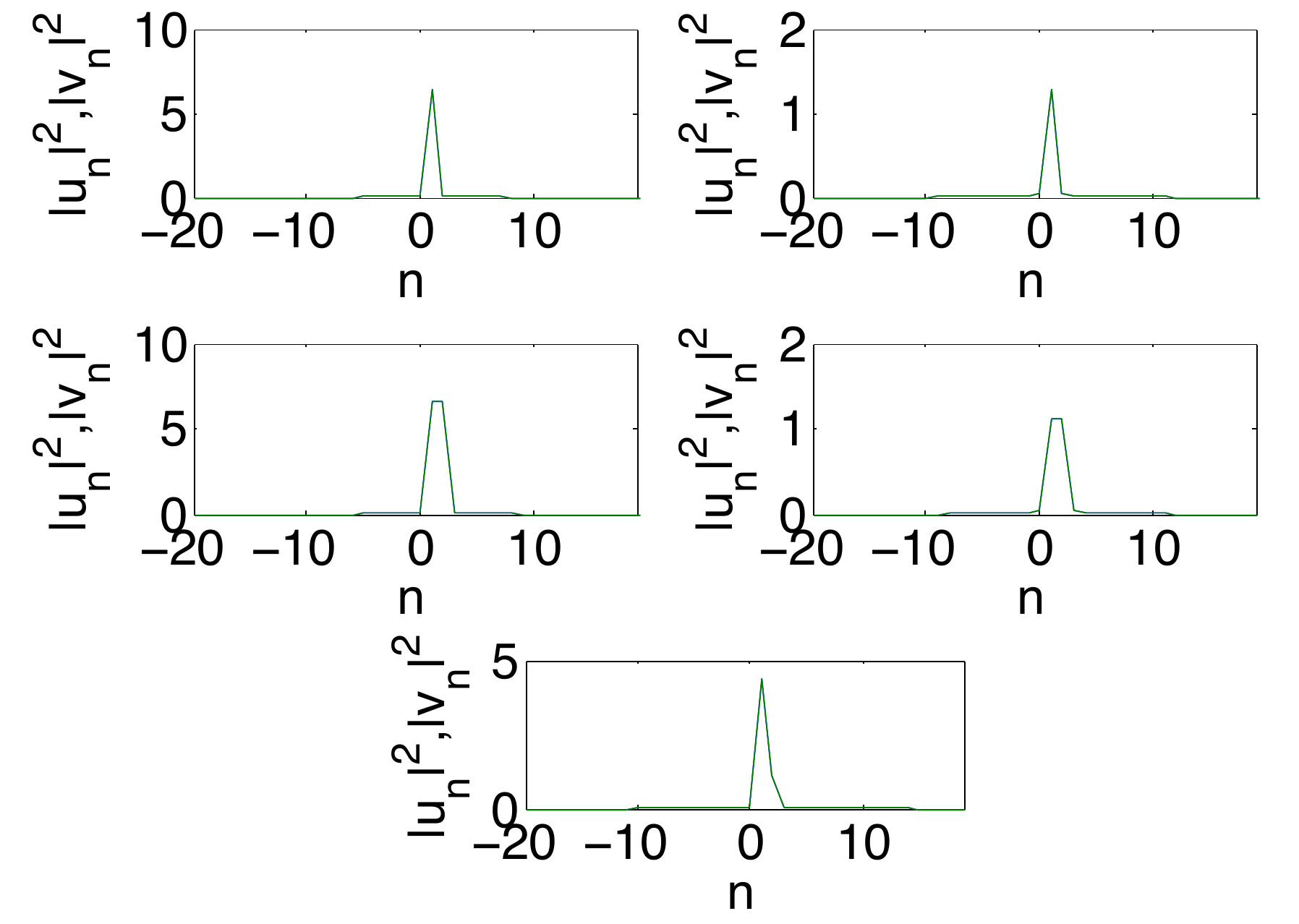} }
\caption{Profiles of the discrete solitons at $C=0.4$, for $\protect\sigma %
=-1$. The configurations of the initial $C=0$ solution and parameters follow
the same pattern as in Fig. \protect\ref{s=-1abs}. }
\label{s=-1prof}
\end{figure}

\begin{figure}[tbp]
\centerline{\includegraphics[scale=.55]{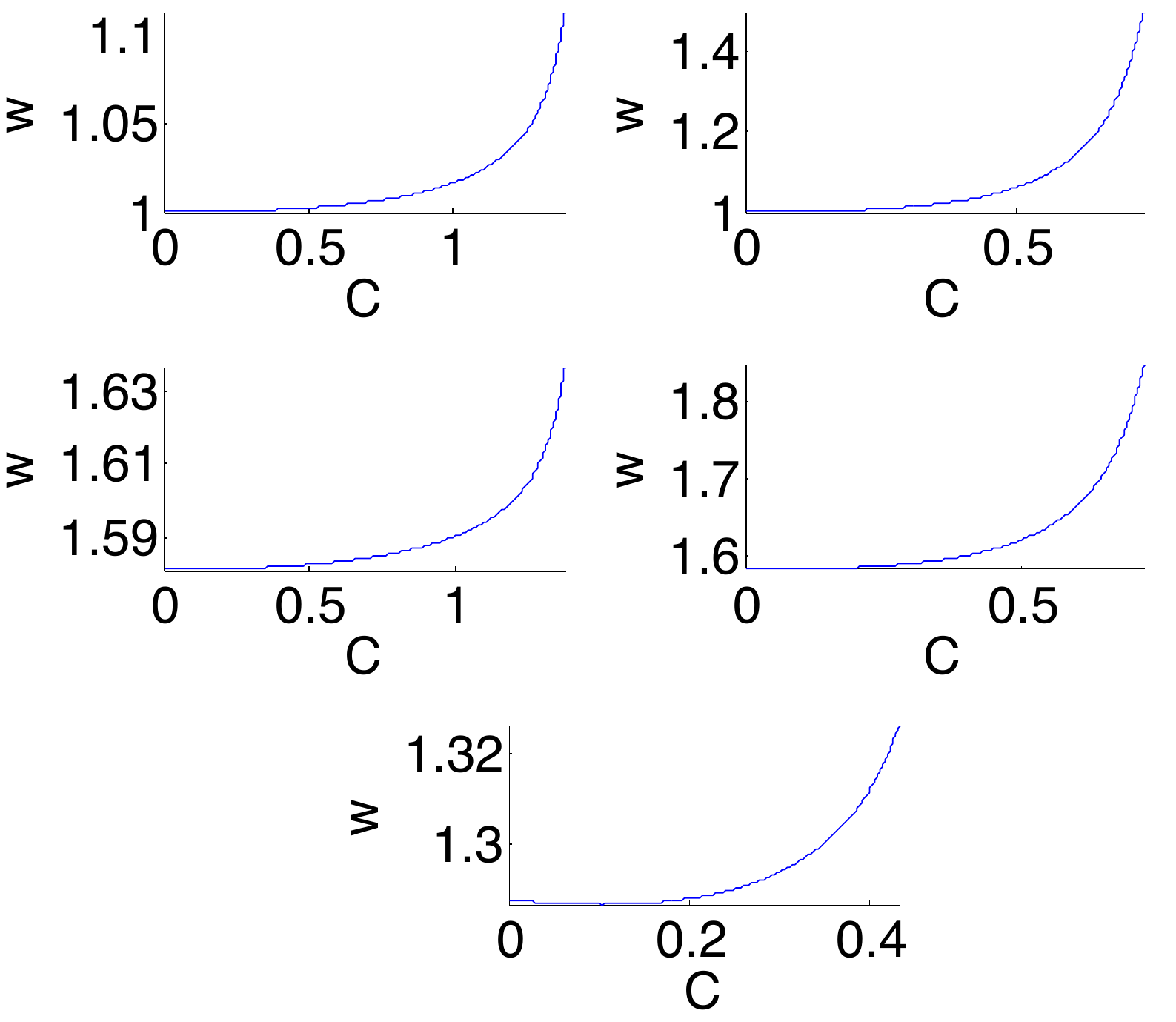} }
\caption{(Color online) The width diagnostic of the discrete solitons,
defined as per Eq. (\protect\ref{w}), corresponding to each of the plots in
Fig. \protect\ref{s=-1abs}. }
\label{s=-1w}
\end{figure}

\begin{figure}[tbp]
\centerline{\includegraphics[scale=.55]{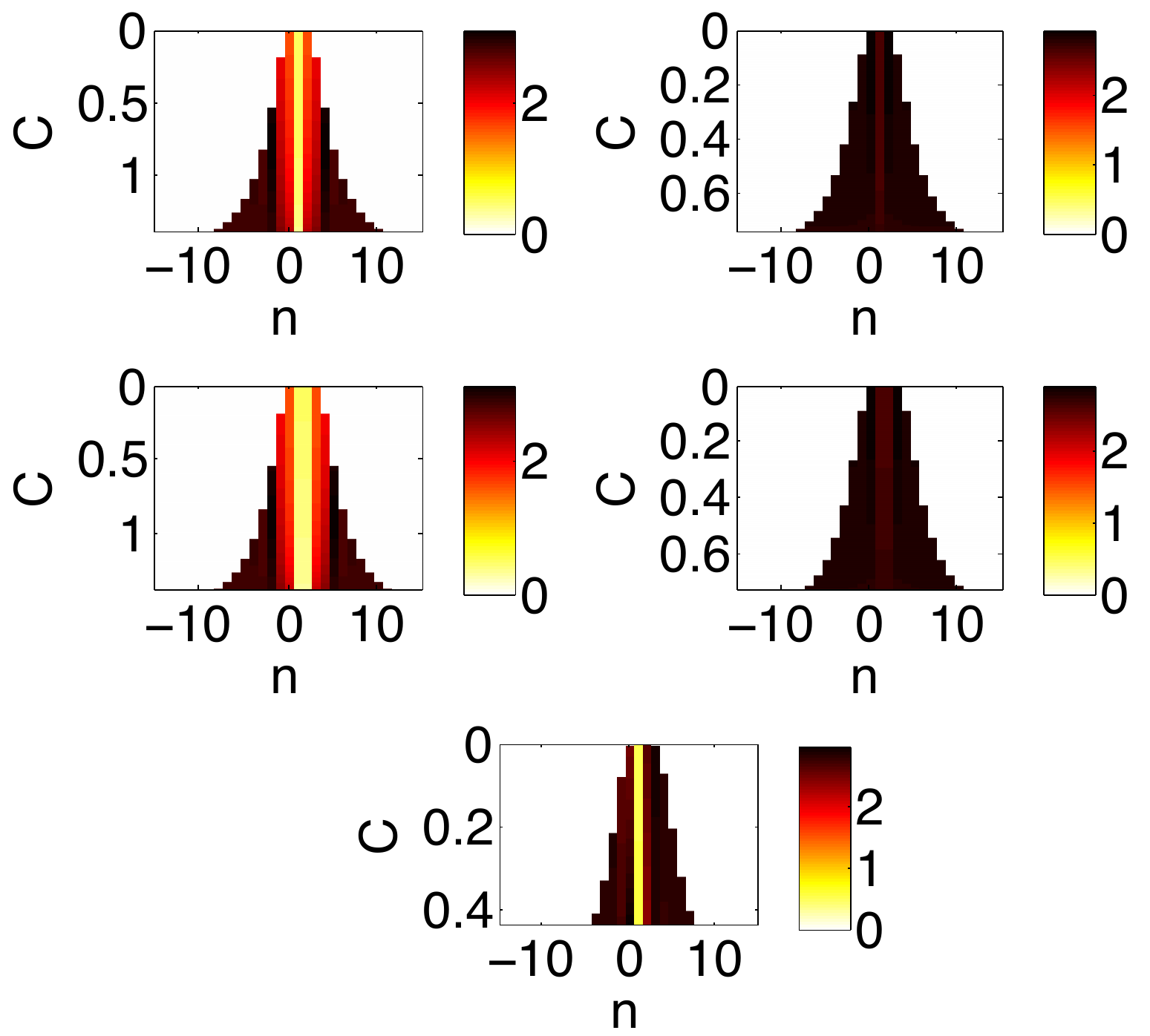} }
\caption{(Color online) The same as Fig. \protect\ref{s=1arg} but for $
\protect\sigma =-1$, i.e., for the discrete solitons presented in Fig.
\protect\ref{s=-1abs}. For the top left and middle left plots, the soliton's
field is nonzero at one or two in-phase rung(s) when $C=0$, and as $C$
increases these rungs remain in-phase, while all the others are
out-of-phase. For the top right and middle right plots, the soliton's field
at $C=0$ is nonzero and out-of-phase at one or two central rungs, and all
rungs remain out-of-phase with the increase of $C$. In the bottom plot, only
the $n=1$ rung remains in-phase, while all others are out-of-phase. }
\label{s=-1arg}
\end{figure}

\begin{figure}[tbp]
\centerline{\includegraphics[scale=.55]{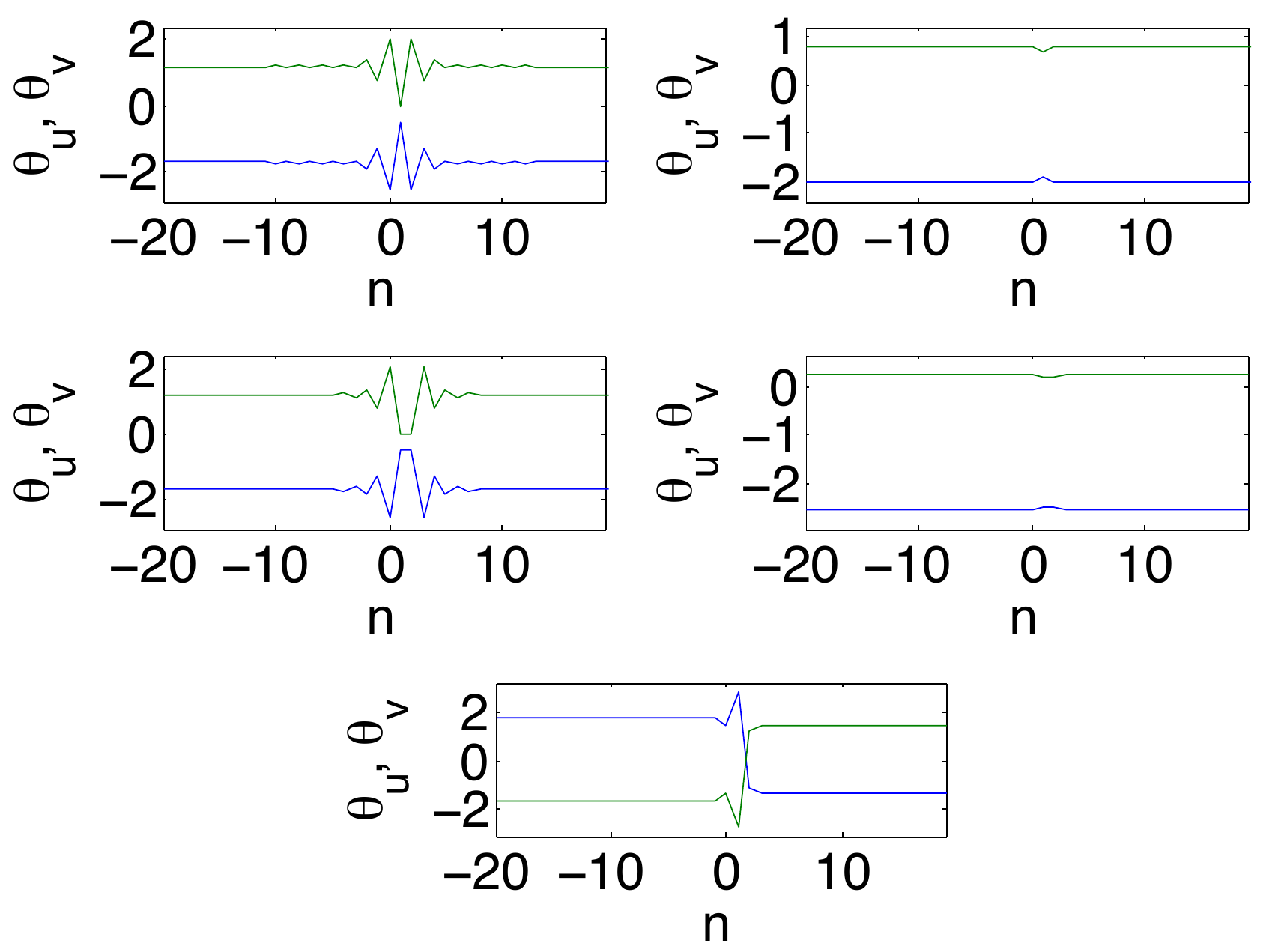} }
\caption{The phase of each rung, $\theta_u := \mathrm{arg}(u_n)$ (blue) and
$\theta_v := \mathrm{arg}(v_n)$ (green) in the interval $(-\pi,\pi]$,
plotted as a function of the spatial variable $n$ for the fixed parameter values $C = 0.4, \gamma = 1$ 
and $\sigma = -1$.  That is, the $u_n, v_n$ solutions are the same as those whole modulus is plotted in 
Fig. \ref{s=-1prof}.
 }
\label{s=-1arguv}
\end{figure}

Figures \ref{s=-1abs}, \ref{s=-1prof}, \ref{s=-1w}, \ref{s=-1arg}, \ref{s=-1arguv} are
similar to their counterparts \ref{s=1abs}, \ref{s=1prof}, \ref{s=1w}, \ref%
{s=1arg}, \ref{s=1arguv}, respectively, but with the parameters taken as per Eq. %
\eqref{pars=-1} instead of Eq. \eqref{par}. Comparing Figs. \ref{s=1arg} and %
\ref{s=-1arg} shows that $\sigma =+1$ favors the solutions with in-phase
rungs as $C$ increases, while $\sigma =-1$ favors the out-of-phase rungs. In
other words, the progressively expanding soliton keeps the in- and
out-of-phase structures, in the case of the self-focusing ($\sigma =+1$) and
defocusing ($\sigma =-1$) onsite nonlinearity, respectively, in agreement
with the well-known principle that discrete solitons feature a staggered
pattern in the case of the self-defocusing \cite{Panos}. Also, according to
Eq. (\ref{rung_rel}), for $\sigma =-1$ the asymmetry of the mixed-phase
solution is switched in comparison to the $\sigma =+1$ case, lending the
in-phase rung a larger magnitude of the fields than in the out-of-phase one.

It is relevant to stress that the discrete solitons seeded in the ACL by
double rungs feature a \textit{bi-dimer} structure which does not carry 
a topological charge \cite{SAM}.
I.e., the solitons cannot take the form of vortices, according to our numerical
results, 
%and according 
%to the argument that a cyclic power flux, which defines
%vortices, cannot be supported by a ring composed of the alternating gain and
%two loss poles 
(contrary to what is the case, e.g.,
for a ring containing a single $\mathcal{PT}$-symmetric
dipole \cite{Leykam}).

\subsection{Stability of the discrete solitons}

\begin{figure}[tbp]
\centerline{\includegraphics[scale=.45]{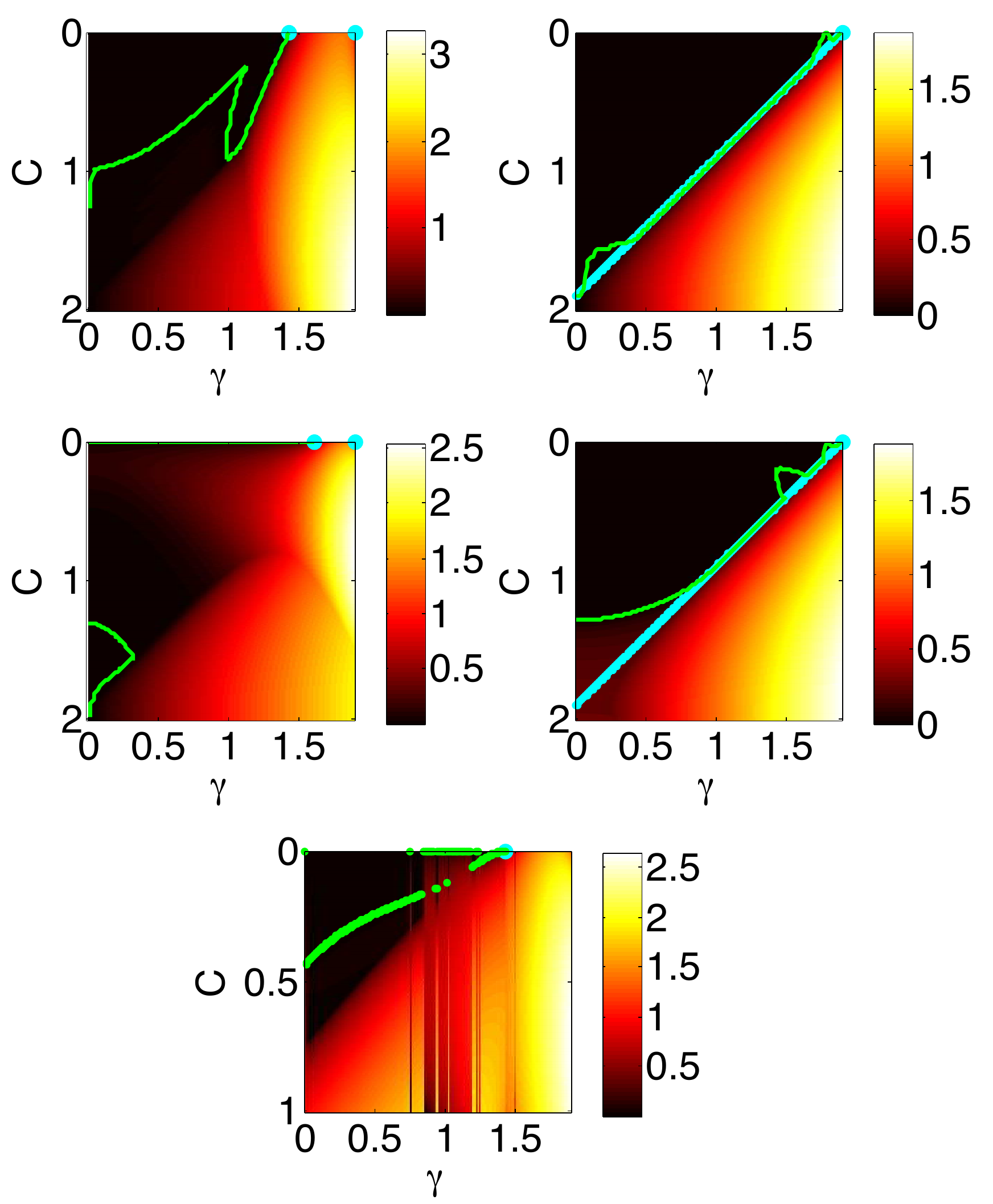} } \vspace{-.2in}
\caption{(Color online) The largest instability growth rate, $\max (\mathrm{%
Re}(i\protect\omega ))$, determined by matrix $M$ in Eq. (\protect\ref{lineq}),
for parameter values following the same pattern as in Fig. \protect\ref%
{s=1abs}, except that here $\protect\gamma $ varies along the horizontal
axis. In the top two plots in the right column, the cyan line represents the
analytically predicted critical value, $\protect\gamma _{\mathrm{cr}%
}^{(1)}(C)=\protect\kappa -C$; this line originates from the cyan dot in the
corner at $\protect\gamma _{\mathrm{cr}}^{(1)}(0)=\protect\kappa $, see Eq. (%
\protect\ref{cr}). If an in-phase excited rung is present, then the second
cyan dot is located at $\protect\gamma _{\mathrm{cr}}^{(2)}=\protect\sqrt{%
\protect\kappa ^{2}-\Lambda ^{2}/4}$, in accordance with Eq. (\protect\ref%
{cr2}). Green lines indicate stability boundaries, between the dark region
corresponding to stability [or very weak instability, with $\max (\mathrm{Re}%
(i\protect\omega ))<10^{-3}$], and the bright region corresponding to the
instability.}
\label{s=1mxRE}
\end{figure}

\begin{figure}[tbp]
\centerline{\includegraphics[scale=.45]{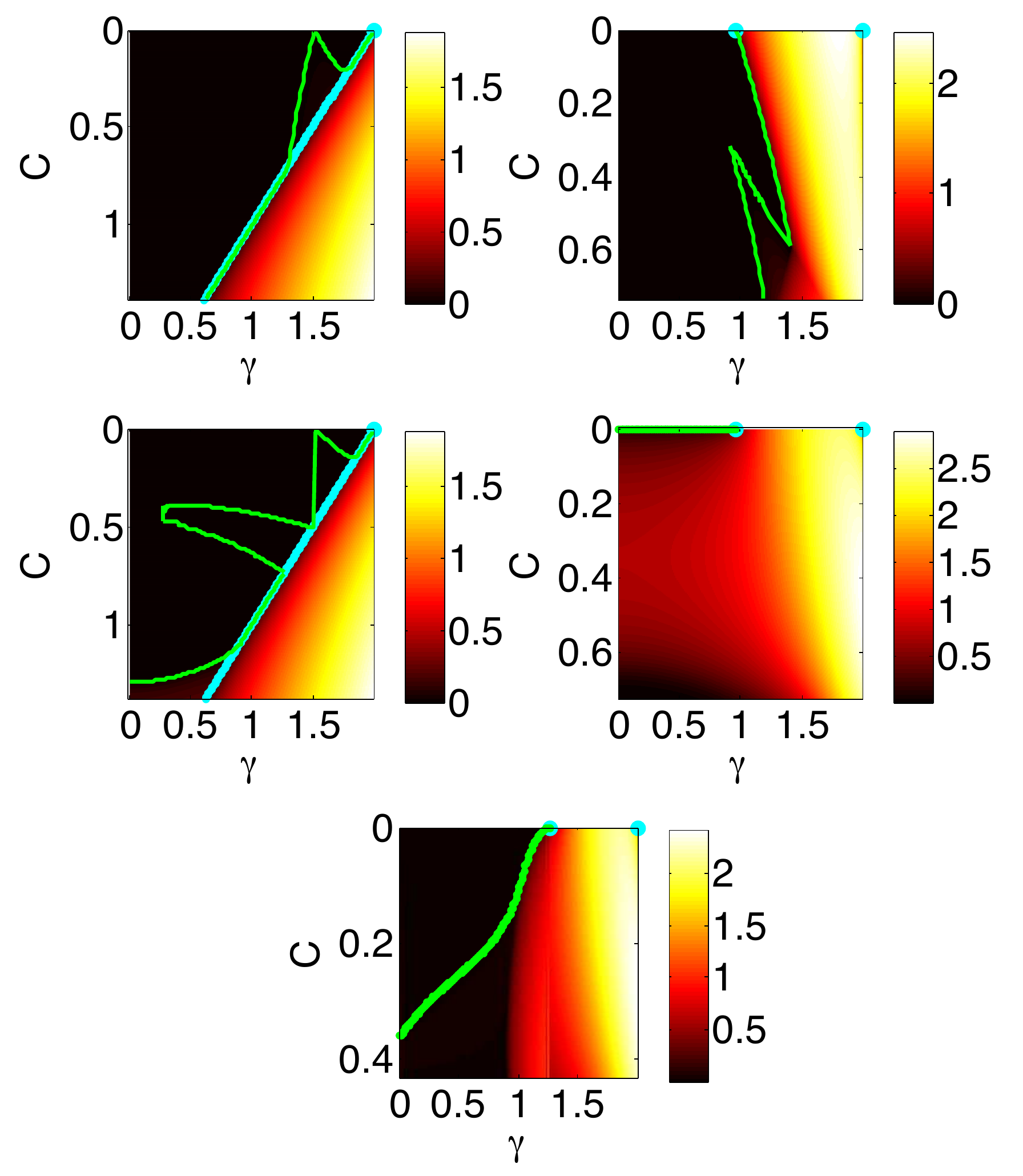} } \vspace{-.2in}
\caption{(Color online) The same as in Fig. \protect\ref{s=1mxRE}, but for
parameter values from Fig. \protect\ref{s=-1abs}. Here the cyan line is
drawn on the top two plots in the left column, and the second cyan dot on
the $C=0$ axis appears only in the case where the out-of-phase excited rung
is present at $C=0$. }
\label{s=-1mxRE}
\end{figure}

Figure \ref{s=1mxRE} shows two-parameter stability diagrams for the solitons
by plotting the largest instability growth rate (if different from zero), $%
\max (\mathrm{Re}(i\omega ))$, as a function of $C$ and $\gamma $ for parameter
values taken per Eq. \eqref{par}, and Fig. \ref{s=-1mxRE} shows the same as
per Eq. \eqref{pars=-1}. The respective stability boundaries are shown by
green lines (white, in the-black-and-white version of the figures). Some
comments are relevant here. Recall that Eqs.~(\ref{cr}) and (\ref{cr2})
impose stability limitations, respectively, from the point of view of the
zero-background solution in the former case, and the single-site excitation
in the latter case. The former background stability condition indicates that
the line of $\gamma =\kappa -C$ (parallel to the antidiagonal cyan line in
Fig. \ref{s=1mxRE}) poses an upper bound on the potential stability of any
excitation, as it is the condition for the stability of the zero background,
on top of which any solitary wave is constructed. It can be seen in both
Figs. \ref{s=1mxRE} and \ref{s=-1mxRE}, especially in the right panels of
the former and left panels of the latter [where the instability defined by
Eq. (\ref{cr2}) is less relevant], that the background-instability threshold
given by Eq. (\ref{cr}) is an essential stability boundary for the family of
the discrete solitons. Of course, additional instabilities due to the
localized core part of the solution are possible too, and, as observed in
these panels, they somewhat deform the resultant stability region. The
additional instabilities stemming from the excited in-phase rungs in Fig. %
\ref{s=1mxRE}, and their out-of-phase counterparts in Fig. \ref{s=-1mxRE},
are separately observed in the left panels of the former figure and right
panels of the latter one. Given that this critical point was found in the
framework of the ACL, it features no $C$ dependence, but it clearly 
contributes to delimiting the stability boundaries of the discrete solitons;
sometimes, this effect is fairly dramatic, as in the middle-row left and
right panels of Figs. \ref{s=1mxRE} and \ref{s=-1mxRE}, respectively, i.e.,
the two-site, same-phase excitations may be susceptible to this instability
mechanism. Although the precise stability thresholds may be fairly complex,
arising from the interplay of localized and extended modes in the nonlinear
ladder system, a general conclusion is that the above-mentioned
instabilities play a critical role for the stability of the localized states
in this system (see also the discussion below). Another essential conclusion
is that the higher the coupling ($C$), the less robust the corresponding
solutions are likely to be, the destabilization caused by the increase of $C$
being sometimes fairly dramatic. %show that
%the critical values $\gamma _{\mathrm{cr}}^{(1)}(C)$, $\gamma _{\mathrm{cr}%
%}^{(2)}(C)$ [see Eqs. (\ref{cr}) and (\ref{cr2}) decrease as $C$ increases.
%Note that for $C>0$ the first critical value follows Eq. (\ref{cr}).

\begin{figure}[tbp]
\caption{(Color online) Stability eigenvalues $i\protect\omega $ in the complex
plane, for parameters chosen in accordance with the top-most left panel of
Fig. \protect\ref{s=1mxRE} with $\protect\gamma =0.5$. For $C=0$, in the top
left plot we show the agreement of the numerically found eigenvalues (blue
circles) with results produced by Eqs. \eqref{nexeig} (green filled circles) and
\eqref{exeig} (red filled). For $C=0.3$, in the top right panel we show
that the eigenvalues associated with the zero solution indeed lie within the
predicted intervals \eqref{intervals}, the boundaries of which are shown by
dashed lines. Next, for $C=1$, in the bottom left plot we observe that
values of $i\protect\omega $ associated with the excited state have previously
(at smaller $C$) merged with the dashed intervals, and now an unstable
quartet has emerged from the axis. For $C=1.5$, in the bottom right panel
the critical point corresponding to Eq. (\protect\ref{cr}) is represented,
where unstable eigenvalues emerge from the axis at the values of $\pm
(\Lambda +C)$, as the intervals in Eq. \eqref{intervals} merge. Comparing
plots in the bottom row, we conclude that the critical point of the latter
type gives rise, in general, to a stronger instability than the former one. }
\label{s=1eigs}\centerline{\includegraphics[scale=.5]{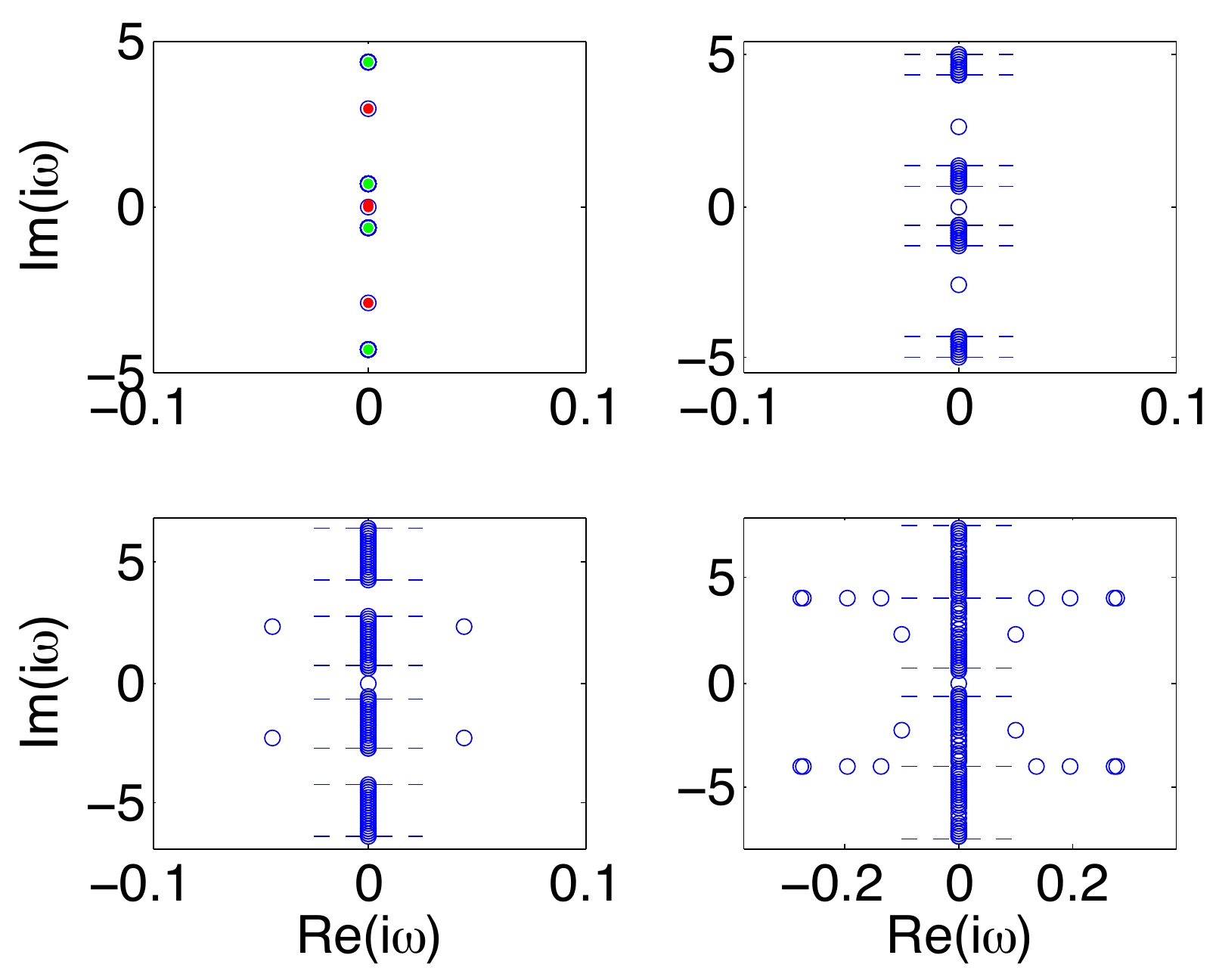} }
\end{figure}

\begin{figure}[tbp]
\caption{ (Color online) The same as in Fig. \protect\ref{s=1eigs}, but for
parameters chosen in accordance with the top right panel of Fig. \protect\ref%
{s=-1mxRE}, with $\protect\gamma =1.2$. At $C=0$, in the top left plot we
show the agreement of the numerically found eigenvalues (blue circles) with
Eqs. \eqref{nexeig} (green filled circles) and \eqref{exeig} (red filled). At $%
C=0.275$, in the top right we see that eigenvalues associated to the red x's
have moved inward towards zero. Next, for $C=0.3$, in the bottom left panel
we observe that, after merging with zero, the eigenvalues now emerge from
zero on the imaginary axis. Finally, at $C=0.7$ in the bottom right panel,
we observe that, after the eigenvalues merge with the dashed-line intervals,
an unstable quartet emerges from the axis. }
\label{s=-1eigs}\centerline{\includegraphics[scale=.5]{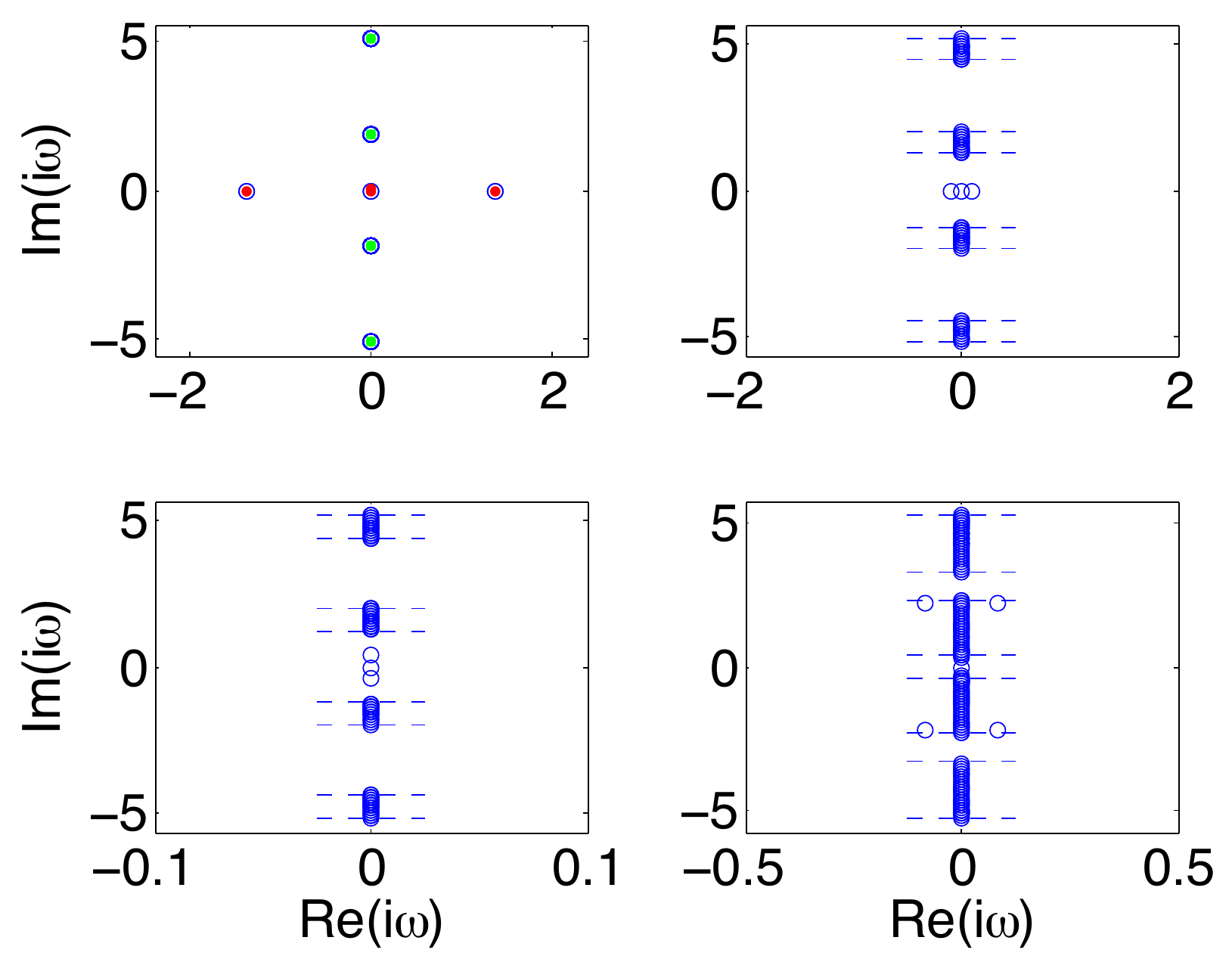} }
\end{figure}

The values of $i\omega $ whose maximum real part is represented in Figs. \ref
{s=1mxRE} and \ref{s=-1mxRE} were computed with the help of an appropriate
numerical eigenvalue solver.
At $C=0$, the eigenvalues agree with Eqs. \eqref{nexeig} and \eqref{exeig}.
As shown in Figs. \ref{s=1eigs} and \ref{s=-1eigs}, following the variation
of $C$ and $\gamma $, eigenvalues \eqref{nexeig}, associated with the empty
(zero-value) sites, vary in accordance with the prediction of Eq. (\ref
{zernu}), and eigenvalues \eqref{exeig}, associated with excited rungs, also
shift in the complex plane upon variation of $C, \gamma$.  In the case of the 
mixed-phase solutions with asymmetric amplitude (seen in the bottom-most plot of Fig. \ref{s=1prof}), there is a stable region for low values of the parameters $C$ and $\gamma$.  For larger values of $\gamma$ there are parametric intervals (across $C>0$ for fixed $\gamma $) in which discrete solitons with phase profiles different from those
initialized in the ACL of $C=0$ have been identified; see the bottom two plots in Fig. \ref{s=1arguv}. 
These distinct branches of the unstable solutions give rise to \textquotedblleft streaks"
observed in the bottom middle panel of Fig. \ref{s=1mxRE}. 
The amplitude profiles of such
alternate solutions are similar to those shown in the bottom plot of Fig. \ref{s=1prof}, and
the gain in width function defined in (\ref{w}) as a
function of $C$ is similar to the examples shown in the bottom plots of Figs. \ref{s=1abs}, \ref%
{s=1w}. Mechanisms by which solutions become unstable for these alternate
solutions are outlined below.

The most obvious type of the instability is associated with initializing a
solution at $C=0$ from a single unstable rung, i.e., at $\gamma >\gamma _{%
\mathrm{cr}}^{(2)}(C=0)$ in (\ref{cr2}). Eigenvalues for this type of the
instability are shown in the top two panels of Fig. \ref{s=-1eigs}. There
are three other scenarios of destabilization of the discrete solitons with
the increase of $C$, each corresponding to a particular type of a critical
point (transition to the instability). These transitions are demonstrated in
Figs.~\ref{s=1eigs}-\ref{s=-1eigs}. The first type occurs when eigenvalue $%
\omega $ associated with an excited rung collides with one of the intervals in
Eq. \eqref{intervals}. This weak instability generates an eigenfrequency
quartet and is represented in Figs. \ref{s=1mxRE} and \ref{s=-1mxRE}, where
the green boundary deviates (as $C$ increases from $0$) from the threshold
given by Eq.~(\ref{cr2}). Figures \ref{s=1eigs} and \ref{s=-1eigs}
illustrate this type of transition in more detail by plotting the
eigenvalues directly in the complex plane.

The second type of the transition occurs when the intervals in Eq. %
\eqref{intervals} come to overlap at $\gamma =\gamma _{\mathrm{cr}}^{(1)}(C)$%
, see Eq. (\ref{cr}). This is the background instability at empty sites, as
shown in Figs. \ref{s=1mxRE} and \ref{s=-1mxRE} by bright spots originating
from the corners of the diagrams, where $\gamma =\kappa =\gamma _{\mathrm{cr}%
}^{(1)}(C=0)$. A more detailed plot of these eigenvalues and the
corresponding collisions in the complex eigenvalue plane is displayed in
Fig. \ref{s=1eigs}.

%The first two types of instabilities are overpowered by the third type,
The third type of the instability onset occurs for essentially all values of
$C$ in the case of two in-phase rungs at $\sigma >0$, or two out-of-phase
ones at $\sigma <0$.
%when $\gamma >\gamma _{\mathrm{cr}}^{(2)}(C)$, see Eq. (\ref%
%{cr2}).
It may be thought of as a localized instability due to the simultaneous
presence of two potentially unstable elements, due to the instability
determined by Eq.~(\ref{cr2}).
%This type is the localized instability associated to the presence of
%an in-phase or out-of-phase rung if $\sigma >0$ or $\sigma <0$,
%respectively. For more details we refer back to Section \ref{solC=0}, where
%we discussed this type for $C=0$.
At $C>0$, it is seen as the bright spots in Figs. \ref{s=1mxRE} and \ref%
{s=-1mxRE} originating from $\gamma _{\mathrm{cr}}^{(2)}(C=0)=\sqrt{\kappa
^{2}-\Lambda ^{2}/4}$. The eigenvalues emerge from the corresponding zero
eigenvalues at $C=0$. That is, in the middle-row left plot of Fig. \ref%
{s=1mxRE} at $C=0$ for $\gamma <\gamma _{\mathrm{\ cr}}^{(2)}(C=0)$ there
are four zero eigenvalues; as $C$ increases, two of the four eigenvalues
move from zero onto the real axis in the complex plane. A similar effect is
observed at $\gamma <\gamma _{\mathrm{\ cr}}^{(2)}(C=0)$ in the middle-row
right plot of Fig. \ref{s=-1mxRE}.

Finally, it is worth making one more observation in connection, e.g., to
Fig. \ref{s=-1eigs} and the associated jagged lines in the top right panel
of Fig.~\ref{s=-1mxRE}. Notice that, as $C$ increases, initial stabilization
of the mode unstable due to the criterion given by Eq.~(\ref{cr2}) takes
place, but then a collision with the continuous spectrum on the imaginary
axis provides destabilization anew. It is this cascade of events that
accounts for the jaggedness of the curve in the top right of Fig.~\ref%
{s=-1mxRE} and in similar occurrences (e.g., in the top left plot of Fig. %
\ref{s=1mxRE}). We add this explanation to the set of possible instabilities
discussed above, to explain the complex form of the stability boundaries
featured by our two-dimensional plots.

\subsection{The evolution of discrete solitons}

\label{prop}

\begin{figure}[tbp]
\centerline{\includegraphics[scale=.5]{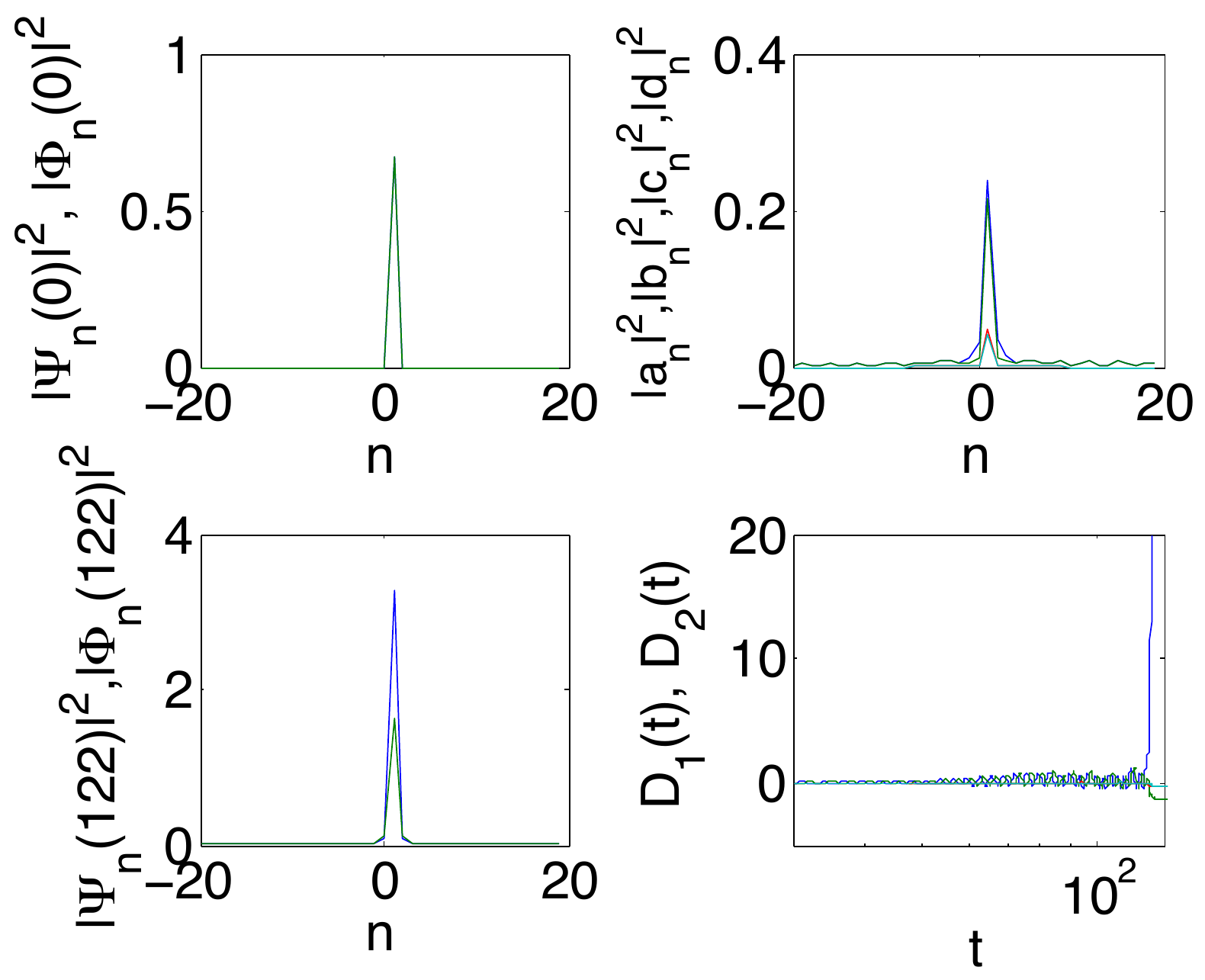} }
\caption{(Color online) The evolution of the soliton whose instability is
predicted in the top-most left panel of Fig. \protect\ref{s=1mxRE} for $C=1$
and $\protect\gamma =0.5$. The complex plane of all the eigenvalues for this
solution is shown in the bottom left plot of Fig. \protect\ref{s=1eigs}. The
top right plot shows the squared absolute values of perturbation amplitudes $%
a_{n},c_{n}$ (higher amplitudes) and $b_{n},d_{n}$ (lower amplitudes),
defined in Eq. \eqref{pert}. The top left plot shows the solution at $t=0$,
and the bottom left plot shows the solution at $t=122$ with $|\Psi
_{n}(t=122)|^{2}$ in blue and $|\Phi _{n}(t=122)|^{2}$ in green. In the
course of the evolution, the soliton maintains its shape, while the
amplitude at the central rung ($n=1$) grows with oscillations; the growth on
the gain side, associated to $\Psi _{n}$, is ultimately dominant. Quantities
$D_{1}(t)\equiv |\Psi _{1}(t)|^{2}-|\Psi _{1}(0)|^{2}$ and $D_{2}(t)\equiv
|\Phi _{1}(t)|^{2}-|\Phi _{1}(0)|^{2}$ are shown in the bottom right plot,
in order to better demonstrate the growing oscillations. }
\label{type1prop}
\end{figure}

\begin{figure}[tbp]
\centerline{\includegraphics[scale=.5]{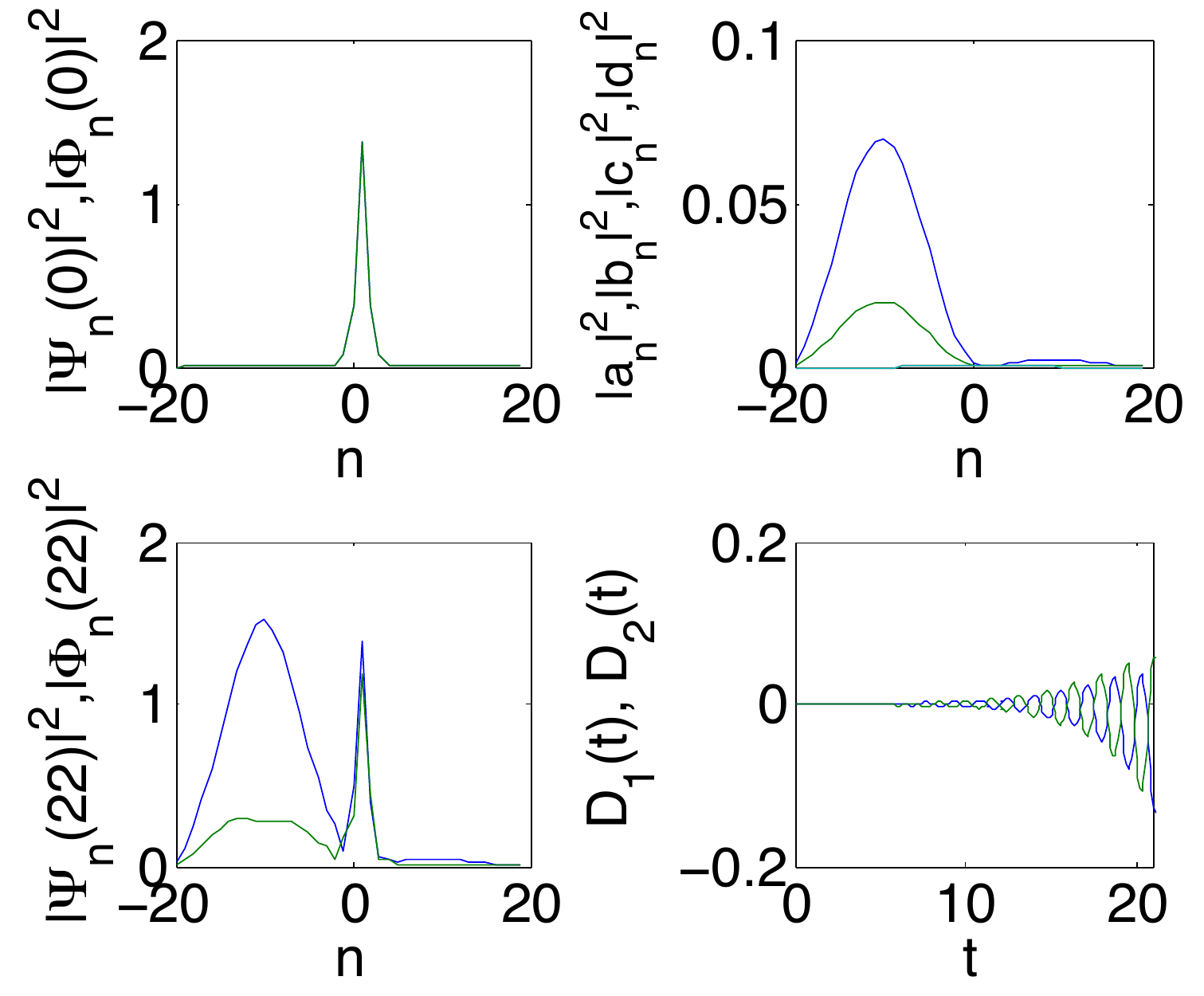} }
\caption{(Color online) The evolution of the discrete soliton whose
instability is predicted is in the top-most left panel of Fig. \protect\ref%
{s=1mxRE} for $C=1.5$ and $\protect\gamma =0.5$. The complex plane of all
the eigenvalues for this solution is shown in the bottom right plot of Fig.
\protect\ref{s=1eigs}. The top right plot has the same meaning as in Fig.
\protect\ref{type1prop}, where here $|b_{n}|^{2},|d_{n}|^{2}$ are mostly
zero while $|a_{n}|^{2},|c_{n}|^{2}$ have nonzero amplitudes. In the course
of the evolution the soliton does not maintain its shape. In particular, the
solution profiles at $t=22$ are shown in the left panel at the bottom
bearing the apparent signature of the delocalized, unstable eigenmode; the
delocalization is stronger in the $\Psi $ solution associated with gain.
Similar to Fig. \protect\ref{type1prop} we plot $D_{1}(t),D_{2}(t)$ in the
bottom right plot. This plot shows that the central node experiences
oscillations similar to Fig. \protect\ref{type1prop} but the oscillatory
effect is dominated by the delocalization seen in the bottom left plot,
which grows and exceeds past the shorter peaks in the center.}
\label{type2prop}
\end{figure}

\begin{figure}[tbp]
\centerline{\includegraphics[scale=.5]{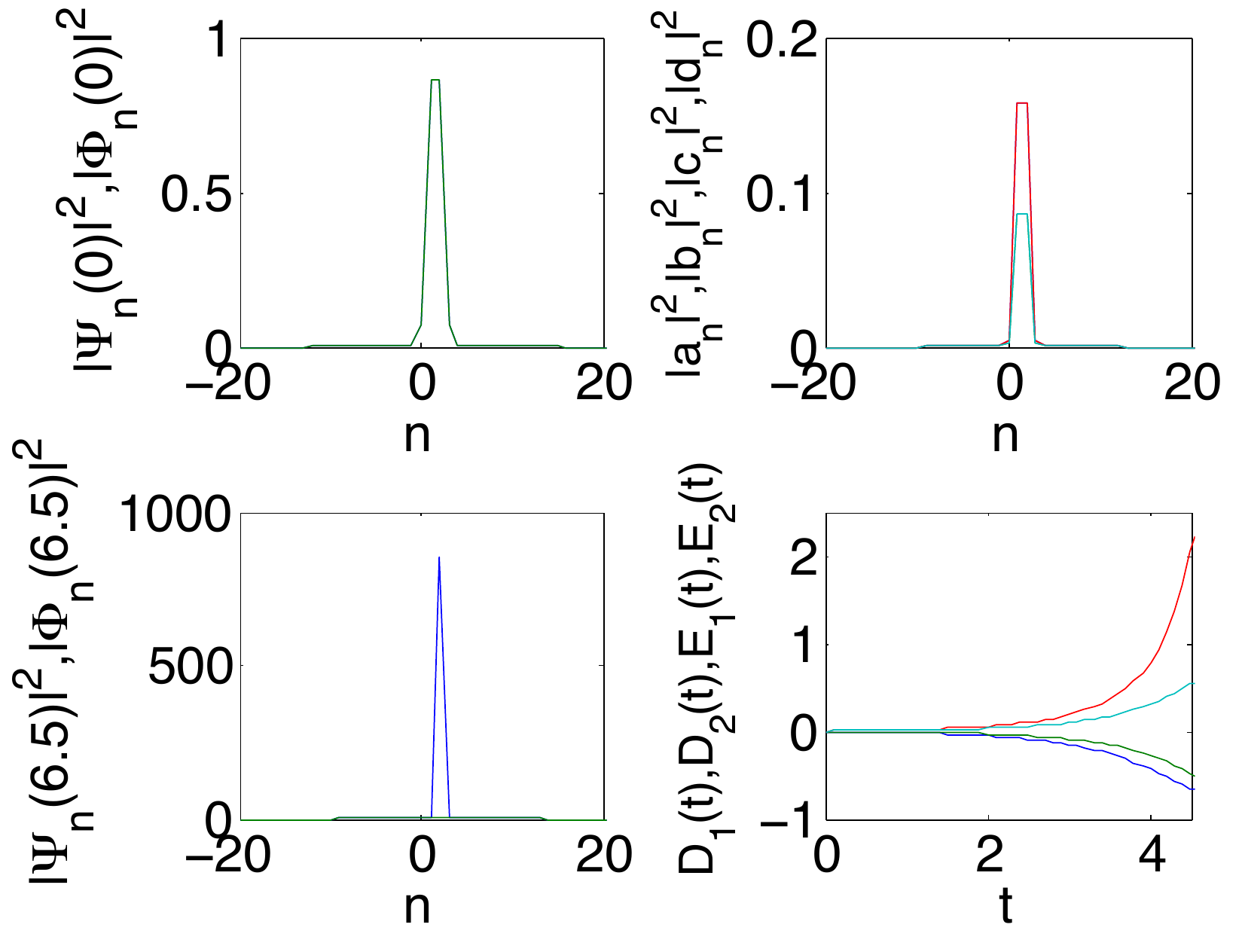} }
\caption{(Color online) The evolution of the solitary wave whose instability
is predicted is in the middle left panel of Fig. \protect\ref{s=1mxRE} for $%
C=0.5$ and $\protect\gamma =1.5$. The complex plane including all the
eigenvalues for this solution is similar to the top right plot in Fig.
\protect\ref{s=-1eigs}. The top right plot here shows the squared absolute
values of perturbation amplitudes $a_{n},b_{n}$ (lower amplitudes) and $%
c_{n},d_{n}$ (higher amplitudes) defined in Eq. \eqref{pert}. The top left
plot shows the solution at $t=0$, and the bottom left plot shows the
solution at $t=6.5$ with $|\Psi_n(6.5)|^2$ in blue and $|\Phi_n(6.5)|^2$ in
green. In the course of the evolution, the soliton maintains its shape,
while the amplitude at the rung $n=2$ grows with weak oscillations and with
higher growth on the $\Psi$ side associated with gain. Quantities $%
D_{1}(t)\equiv |\Psi _{1}(t)|^{2}-|\Psi _{1}(0)|^{2}$ (blue), $%
D_{2}(t)\equiv |\Phi _{1}(t)|^{2}-|\Phi _{1}(0)|^{2}$ (green), $%
E_{1}(t)\equiv |\Psi _{2}(t)|^{2}-|\Psi _{2}(0)|^{2}$ (cyan) and $%
E_{2}(t)\equiv |\Phi _{2}(t)|^{2}-|\Phi _{2}(0)|^{2}$ (red) are shown in the
bottom right plot to highlight the growth. }
\label{type3prop}
\end{figure}

To verify the above predictions for the stability of the discrete solitons,
we simulated evolution of the perturbed solutions in the framework of Eq. %
\eqref{tdnls} by means of the standard Runge-Kutta fourth-order integration
scheme. In Figs. \ref{type1prop}, \ref{type2prop}, and \ref{type3prop} we
display examples of the evolution of each of the three instability types
which were identified in Section \ref{Cneq0}.

For the first type, when the instability arises from the collision of
eigenvalues associated with the excited and empty rungs, the corresponding
unstable eigenmode arises in the form of a quartet of eigenfrequencies. In
Fig. \ref{type1prop} we demonstrate that this instability leads to the
growth of the solution amplitudes and oscillations at the central rung. The
corresponding (chiefly localized, although with a weakly decaying tail)
instability eigenvectors are shown in the top panels of the figure, while
the bottom panels show how the initial conditions evolve in time through the
oscillatory growth, in accordance with the presence of the unstable complex
eigenfrequencies.

For the second type of the instability, which arises from the collision of
eigenvalues in intervals \eqref{intervals}, which are all associated with
empty rungs, the corresponding unstable eigenmode is delocalized. It is
shown in Fig. \ref{type2prop} that the corresponding unstable soliton does
not preserve its shape. Instead, the instability causes delocalization of
the solution, which acquires a tail reminiscent of the spatial profile of
the corresponding unstable eigenvector.

%\begin{figure}[tbp]
%\centerline{\includegraphics[scale=.5]{s=-1C=_2prop_new.pdf} }
%\caption{(Color online) The evolution of the soliton whose instability is predicted is in the top-most right panel of Fig. \protect\ref{s=-1mxRE} for $C=0.2$ and $\protect\gamma =1.2$. The complex plane including all the eigenvalues for this solution is similar to the top right plot in Fig. \protect\ref{s=-1eigs}. The top right plot has the same meaning as in Figs. \protect\ref{type1prop} and \protect\ref{type2prop}. In the course of the evolution, the soliton maintains its shape, while the amplitude at the central rung, $n=1$, grows with weak oscillations and with higher growth on the $\Psi$ side associated with gain. Quantities $D_{1}(t)\equiv |\Psi _{1}(t)|^{2}-|\Psi_{1}(0)|^{2}$ and $D_{2}(t)\equiv |\Phi _{1}(t)|^{2}-|\Phi _{1}(0)|^{2}$ are shown in the bottom right plot to highlight the growth. }
%\label{type3prop}
%\end{figure}

Lastly, the third type of the instability is shown
%of instability associated with the criterion
%of Eq. (\ref{cr2}) leads to an indefinite growth for
%$\gamma >\gamma _{\mathrm{cr}}^{(2)}$ (for $C=0$, and accordingly
%for modified values of $\gamma _{\mathrm{cr}}^{(2)}$ as $C$ grows).
%This is consonant to what is known from the single dimer case
%e.g. in~\cite{PT1,Zezyulin}.  We show
in Fig. \ref{type3prop}. %, which concerns the case
% of two in-phase excited rungs in the focusing realm of $\sigma=1$.
It displays the case of two excited in-phase rungs at $\sigma =1$. Other
examples of the same type are similar
%, such as the case of $\sigma=1$ with one in-phase excited rung,
-- e.g., with two out-of-phase excited rungs at $\sigma =-1$. The
instability has a localized manifestation with the amplitudes growing at the
gain nodes of each rung and decaying at the loss ones.
% the third type of the instability arises at $\gamma >\gamma _{%
%\mathrm{cr}}^{(2)}$ [see Eq. (\ref{cr2})] due to the presence of an in- or
%out-of-phase excited rung, at $\sigma >0$ and $\sigma <0$, respectively.
%Figure \ref{type3prop} demonstrates that this strongest type of the
%instability results in growth of amplitudes at the central rungs while
%maintaining the soliton's shape.

\section{Conclusions}

We have introduced the lattice of the ladder type with staggered pairs of
mutually compensated gain and loss elements at each rung, and the usual
onsite cubic nonlinearity, self-focusing or defocusing. This
nearly-one-dimensional system is the simplest one which features\emph{\
two-dimensional }$\mathcal{PT}$ symmetry. It may be realized in optics as a
waveguide array. We have constructed families of discrete stationary
solitons seeded by a single excited rung, or a pair of adjacent ones, in the
anti-continuum limit of uncoupled rungs. The seed excitations may have the
in-phase or out-of-phase structure in the vertical direction (between the
gain and loss poles). The double seed with the in- and out-of-phase
structures in the two rungs naturally features an asymmetric amplitude
profile. We have identified the stability of the discrete solitons via the
calculation of eigenfrequencies for small perturbations, across the system's
parameter space. A part of the soliton families are found to be dynamically
stable, while unstable solitons exhibit three distinct scenarios of the
evolution. The different scenarios stem, roughly, from interactions of
localized modes with extended ones, from extended modes alone, or from
localized modes alone.

A natural extension of the work may be the consideration of mobility of
kicked discrete solitons in the present ladder system. It may also be
interesting to seek  nonstationary solitons with periodic intrinsic
switching, cf. Ref. \cite{AD}. A challenging perspective is the development
of a 2D extension of the system. Effectively, this would entail adding
further alternating ladder pairs along the transverse direction and
examining 2D discrete configurations. 
It may be relevant in such 2D extensions to
consider different lattice settings that support not only 
solutions in the form of discrete
solitary waves but also ones 
built as discrete vortices, similarly to what has been
earlier done in the DNLS system~\cite{Panos}, and recently in another 2D $%
\mathcal{PT}$-symmetric system \cite{Raymond-OE}.

\end{document}